\documentclass[10pt,journal,twocolumn]{IEEEtran}
\newif\ifCLASSOPTIONromanappendices \CLASSOPTIONromanappendicestrue
\usepackage{a0size}
\usepackage{amssymb}
\usepackage{multicol}
\usepackage[english]{babel}
\usepackage{epsfig}
\usepackage{bm}
\usepackage{amsfonts,color,amsthm,amsmath, lscape}%,graphics}
\usepackage{url}
\usepackage{algpseudocode}

\usepackage[font=footnotesize]{caption}
\usepackage{subcaption}
\usepackage{epstopdf}
\usepackage{cite}
\usepackage{relsize}
\usepackage{hyperref}
\usepackage{fix2col}
\DeclareMathOperator*{\Maximize}{maximize}

\renewcommand{\figurename}{Fig.}
\addto\captionsenglish{\renewcommand{\figurename}{Fig.}}

\usepackage[ruled]{algorithm2e}
\usepackage{graphicx}
\usepackage{textcomp}
\usepackage{xcolor}
\usepackage{optidef}
\def\BibTeX{{\rm B\kern-.05em{\sc i\kern-.025em b}\kern-.08em
    T\kern-.1667em\lower.7ex\hbox{E}\kern-.125emX}}

\newcommand{\bb}{\mathbf{b}}

\newcommand{\bh}{\mathbf{h}}

\newcommand{\bv}{\mathbf{v}}
\newcommand{\bV}{\mathbf{V}}

\newcommand{\bI}{\mathbf{I}}

\newcommand{\bx}{\mathbf{x}}
\newcommand{\bX}{\mathbf{X}}

\newcommand{\bs}{\mathbf{s}}

\newcommand{\by}{\mathbf{y}}
\newcommand{\bz}{\mathbf{z}}

\newcommand{\bH}{\mathbf{H}}

\newcommand{\bq}{\mathbf{q}}

\newcommand{\bW}{\mathbf{W}}

\renewcommand{\frac}{\dfrac}

\newcommand{\Tr}{{\mbox{Tr}}}

\definecolor{myOrange}{rgb}{1,0.5,0}
\definecolor{myGreen}{rgb}{0,0.5,0}

\newcommand{\changer}[1]{{\color{black}#1}}
\newcommand{\changeRR}[1]{{\color{black}#1}}
\newcommand{\editb}[1]{{\color{black}#1}}
\newcommand{\changeNEW}[1]{{\color{black}#1}}

\begin{document}
\title{Deep Learning for Distributed Channel Feedback and Multiuser Precoding in FDD Massive MIMO
}

\author{Foad~Sohrabi,~\IEEEmembership{Member,~IEEE,}
        Kareem~M.~Attiah,~\IEEEmembership{Student~Member,~IEEE,}
        and~Wei~Yu,~\IEEEmembership{Fellow,~IEEE}
\thanks{The authors are with The Edward S.\ Rogers Sr.\ Department of Electrical and Computer Engineering, University of Toronto, Toronto, ON M5S 3G4, Canada (e-mails:\{fsohrabi, kattiah, weiyu\}@ece.utoronto.ca). This work is supported by Huawei Technologies Canada. \changeNEW{The source code for this paper are available at: \url{https://github.com/foadsohrabi/DL-DSC-FDD-Massive-MIMO}}
}}

\maketitle
%%%%%%%%%%%%%%%%%%%%%%%%%%%%%%%%%%%
% Abstract
%%%%%%%%%%%%%%%%%%%%%%%%%%%%%%%%%%%
\begin{abstract}
This paper shows that deep neural network (DNN) can be used for efficient and 
distributed channel estimation, quantization, feedback, and downlink multiuser
precoding for a frequency-division duplex massive multiple-input
multiple-output system in which a base station (BS) serves multiple
mobile users, but with rate-limited feedback from the users to the BS. 
A key observation is that the multiuser
channel estimation and feedback problem can be thought of as a
distributed source coding problem.  In contrast to the traditional approach
where the channel state information (CSI) is estimated and quantized at each
user independently, this paper shows that a joint design of \changeNEW{pilots} and a new DNN
architecture, \changeNEW{which} maps the received pilots directly into feedback bits
at the user side then maps the feedback bits from all the users directly into
the precoding matrix at the BS, can significantly improve the overall performance. 
This paper further proposes robust design strategies with respect to
channel parameters and also \changeNEW{a} generalizable DNN architecture for varying 
number of users and number of feedback bits.  Numerical results show that
the DNN-based approach with short pilot
sequences and very limited feedback overhead can already approach the
performance of conventional linear precoding schemes with full CSI.  

\end{abstract}


%%%%%%%%%%%%%%%%%%%%%%%%%%%%%%%%%%%%
% I) Introduction
%%%%%%%%%%%%%%%%%%%%%%%%%%%%%%%%%%%%
\section{Introduction}
\label{sec:intro}

Machine learning methods, especially deep neural networks (DNNs), have recently 
shown great potential in dealing with complex optimization problems in various
wireless communications settings, e.g., MIMO detection \cite{Wiesel2019},
massive multiple-input multiple-output (MIMO) hybrid precoding
\cite{Huang2019}, constellation design \cite{Foad2019Asilomar,Foad2020ICASSP}, user scheduling
\cite{WeiWilliam2019}, etc. The data-driven approach has an advantage from both
performance and complexity perspectives, especially when 
the models are uncertain and when conventional optimization approaches have
high complexity. 
This paper aims to show the effectiveness of the deep learning framework 
for obtaining optimized feedback and beamforming strategies for the
frequency-division duplex (FDD) massive MIMO system design. The key observation of this paper is that the multiuser 
channel estimation, quantization and feedback problem can be thought of as a
\emph{distributed source coding (DSC)} problem for which the optimal solution is
analytically intractable, but where the machine learning approach can have an important 
advantage. 

Massive MIMO is an indispensable technology for addressing the ever-increasing
demand for data rate in the next generations wireless networks
\cite{andrews2014will}.  The canonical massive MIMO system operates in the
time-division duplex (TDD) mode in which the BS can exploit the uplink-downlink
channel reciprocity via uplink training to acquire channel state information
(CSI) and subsequently use the CSI for downlink precoding \cite{Marzetta2010}.
However, many existing wireless networks operate in FDD mode in which
uplink-downlink channel reciprocity cannot be assumed. For these FDD systems,
the channels need to be estimated in the downlink, then fed back to the BS for
precoding.  The question of how to optimally design such a channel feedback
strategy is crucial for the wide adoption of FDD massive MIMO in wireless
networks \cite{Emil2016myths}.

This paper focuses on a multiuser FDD downlink massive MIMO system in a 
limited-scattering environment, e.g., in millimeter wave (mmWave) band
\cite{pi2011introduction}, where sparsity can be exploited for channel
estimation. But instead of considering channel estimation as a \changeNEW{stand-alone}
module, we consider the end-to-end system including the design of the downlink
pilot, the channel estimation and quantization strategy under limited feedback, 
as well as the design of the downlink precoding matrix at the BS. 
We make a crucial observation that because \changeNEW{the} downlink channel estimation and
quantization take place in a distributed fashion across the users, yet are fed
back to a centralized location at the BS for precoding purpose, the overall
system is akin to a DSC scheme. While traditional wireless system design always
performs independent quantization of each user's channel and never takes the
distributed nature of channel quantization into consideration, this paper shows that:
\begin{itemize}
\item Optimized distributed channel compression strategy can significantly 
outperform the conventional independent CSI estimation and feedback scheme;
\item Joint design of channel feedback and precoding has significant advantage;
\item Deep neural network (DNN) can play a crucial role for the effective 
design of such a joint precoding and distributed channel compression strategy.
\end{itemize}

The information theoretic considerations of DSC have appeared in the literature
since 1970s. The celebrated Slepian-Wolf Theorem shows that optimal lossless
DSC of two or more correlated sources with separate encoders and a joint
decoder can be much more efficient than independent encoding/decoding
\cite{SlepianWolf1973}. Their strategy of using binning to take advantage of the
correlation between sources is further extended to lossy compression by Wyner
and Ziv \cite{WynerZiv1976}. While many results in DSC aim to recover the 
correlated sources at the decoder \cite{Liveris2004,elgamal2011network}, the 
concept of DSC can also be extended to computing a \changeNEW{function of the sources
\cite{Marton1979, Diggavi2020, Nazer2007,Misra2011}.} For example, \cite{Marton1979} characterizes the
rate region for distributed compression of two correlated uniform binary sources
for computing their modulo-2 sum. It is important to note that the benefit of the DSC strategy can
come from exploiting not only the correlation between the sources, but also the fact that a function of the sources 
(rather than the sources themselves) is desired at the decoder. For example, \cite{Diggavi2020} examines the distributed 
source coding problem for classification  \changeNEW{and \cite{Nazer2007,Misra2011} show that}  the benefit of DSC comes from designing separate encoders and joint decoder that exploit a match between 
the quantization scheme and the function to be computed. 
Such benefit exists even for the case where the sources are independent.

\changeNEW{To further illustrate the benefit of DSC for the case where the sources are uncorrelated, consider the following example inspired by a scenario considered in \cite{Kschischang2007}.} Suppose that we have independent $x_1$ and $x_2$ at the two encoders, both uniformly distributed in $[0,1]$. If the decoder needs to reconstruct both $x_1$ and $x_2$, then since they are independent, the best strategy is just the independent and uniform quantization of $x_1$ and $x_2$ separately. But, now consider the case in which the decoder needs to compute $\max(x_1,x_2)$. Clearly, the uniform quantization is no longer optimal. This is because $\max(x_1,x_2)$ has a distribution tilted toward the higher range of $[0,1]$. Here, the optimal quantizer should have more levels in the upper range and fewer levels in the lower range. This is an example where the quantizers at the distributed sources need to be designed to take into account that a \textit{function} of the sources needs to be computed at the decoder, even when the sources are independent.

A key insight of this paper is the recognition that the end-to-end design of a
downlink FDD precoding system can be regarded as a DSC problem of computing a function 
(i.e., the downlink precoding matrix) of independent sources
(i.e., channels) under finite feedback rate constraints. This paper makes the
case that by designing an optimized DSC strategy, we can significantly reduce 
the amount of feedback as compared to the conventional design based on separate CSI estimation and feedback of each user. 
The design of the optimal DSC strategy is, however, a difficult problem in general, even for the
case in which the source distributions are completely known and the blocklength
is large. This motivates us to use the deep learning technique to tackle such an
optimization problem.

The deep learning framework is well suited to tackle the DSC design problem
because of the following. First, as different from the \changeNEW{conventional} communications system design
methodology which optimizes different blocks of a communication system
separately, the deep learning framework can jointly design all the components
for end-to-end performance optimization, making it suitable for
designing \changeNEW{DSC strategies in which} the goal is to compute a function of 
the sources. Second, unlike the classical DSC which requires customized source coding
design for each different scenario, the deep learning framework implicitly
learns the channel distributions in a data-driven fashion in the process of
optimizing the end-to-end communications system, without requiring tractable
mathematical channel models. Third, computation using trained DNN can be highly 
parallelized, so that the computational burden of DNN is manageable. 

%%%%%%%%%%%%%%%%%%%%%%%%%%%%%%%%%%%%
% I-A) Main Contributions
%%%%%%%%%%%%%%%%%%%%%%%%%%%%%%%%%%%%
\subsection{Main Contributions}

This paper shows that the end-to-end downlink precoding design problem for an
FDD massive MIMO system can be viewed as a DSC problem. More specifically, the
channels from the BS to the different users can be considered as the ``sources'',
and the objective of the DSC scheme is to recover a function of these sources,
\changeNEW{namely, an} optimal precoder at the BS that maximizes a system objective, e.g.,
the sum rate of all the users, based on the rate-limited feedback. To tackle this
challenging DSC design problem, this paper proposes a data-driven approach
wherein the channel estimation, distributed compression, feedback at the 
user side, and multiuser precoding at the BS side can be efficiently designed
jointly by training a DNN at each user and a DNN at the BS. In particular, we
propose a novel neural network architecture that encapsulates all the
components of the FDD downlink precoding system. 
By properly training the proposed neural network, we jointly optimize the
downlink pilots and channel estimation process, the uplink channel quantization
and rate-limited feedback scheme at each user, and the multiuser downlink
precoding scheme at the BS for each user.   

The training of the proposed DNN is however also challenging. In the proposed
DNN architecture, the feedback information bits are modeled as the outputs of
binary neurons. Without any adjustments to the conventional
\textit{back-propagation} algorithm, these binary neurons 
would have vanishing gradients during training, which inhibit the
training of network parameters. To overcome this issue, we alternatively
approximate the gradients of the binary layer with a variant of
\textit{straight-through} (ST) estimator \cite{hinton2012videos, bengio2013, chung2016}.
In the numerical experiments, the proposed neural network architecture trained
with the ST approximation technique exhibits outstanding performance,
especially when the channel training and feedback are severely rate limited.
These results confirm that the deep learning framework can indeed be utilized
to obtain near-optimal, yet practical, DSC strategies which are in general hard
to achieve using \changeNEW{conventional} heuristic methods. Further, the results show that the performance gain comes
from the overall joint DSC and precoding design that bypasses individual explicit channel estimation at each user.

While the proposed DNN has excellent performance when the DNN is trained
and tested under the same environment, one of the main questions about any
data-driven approach, including the proposed DNN, is whether or not its
performance is generalizable to unseen system environments. Towards
addressing this \changeNEW{question}, this paper discusses in detail how to make the
proposed architecture generalizable for different system parameters. In
general, we categorize the system parameters into two categories: i) the
system parameters that only change the input distribution of the DNN, e.g., the
number of paths in the sparse channel model and the signal-to-noise ratio (SNR), 
and ii) the system parameters that also change the input/output dimensions of
some layers in the proposed DNN, e.g., the feedback rate limits and the number
of users.  

For the first category of the system parameters, we numerically show
that training the DNN on a wider range of system parameters can help design
more robust systems when prior knowledge about these parameters is not
available. For the second category, however, training a DNN that can operate for different
system dimensions is more challenging. In this regard, this paper proposes to
modify the neural network architecture and its training procedure as a way to
enhance the generalizability of the proposed DNN with respect to the feedback
rate limits as well as the number of users as follows. We propose a novel
two-step training approach to design a common neural network that can operate
over a wide range of feedback rate limits. In the first step of the modified
approach, we train a modified version of the proposed neural network in which
the outputs of the user-side DNNs are soft binary values\footnote{%\changeRR{This paper represents a bit of information 
%by a variable that takes a (bipolar) binary value, meaning that it is either $-1$ or $+1$. Furthermore, 
%we use the term soft binary for variables taking values between $-1$ and $+1$.}
\changeRR{This paper uses the term {\em binary} to denote variables taking values of either -1 or +1, and {\em soft binary} to denote variables taking values between -1 and +1.}} (instead of binary
values). The trained modified network is used to obtain parameters such as the
pilot sequences and the channel estimation scheme. In the second step, we fix 
the user-side DNN, but apply different quantization resolutions to the user-side 
DNN output to account for different feedback rates. For each feedback rate, 
we conduct another round of training in which only the BS-side DNN parameters are trainable. Numerical results show that 
 this two-step training approach performs almost as well as
the architecture that requires one separate trained DNN on
both the BS-side and user-side for each value of the feedback rate. 

\editb{This paper proposes to address the} generalizability of the DNN with respect to the number of users %can be
%achieved 
using a similar idea. In particular, we propose the following two-step training procedure.
First, all different users are assumed to adopt a common set of DNN weights and
biases. We then seek to design those weights and biases together with the
channel estimation pilot sequences using end-to-end training of a \textit{single-user}
system. As the second step, we seek to design the BS-side DNN by training a
$K$-user system in which all the user-side DNNs are fixed to the parameters
obtained from the single-user case. This novel two-step training approach has 
the desirable property that the user-side operations do not depend on the
number of users in the network; we only need to train and store different
BS-side DNNs to handle varying number of users in the
network. Numerical experiments show that for independent and
identically distributed (i.i.d.) channels, the DNN trained using such a two-step
training method can approach the performance of the original DNN \editb{in a typical scenario}. 

%%%%%%%%%%%%%%%%%%%%%%%%%%%%%%%%%%%%
% I-B) Related Work
%%%%%%%%%%%%%%%%%%%%%%%%%%%%%%%%%%%%
\subsection{Related Work}

Most of the exiting schemes for limited feedback multiuser FDD MIMO systems fall into two categories \cite{Love2008}. The first category of works focus on
reducing feedback overhead by exploiting the spatial and/or temporal
correlation of \changeRR{\changer{CSI} \cite{Gao2019, Rao2014, kuo2012,Zhang2020estimation}}. In particular, since
channels in the correlated CSI scenarios can be represented as a function of
uncorrelated sparse vector in some bases (e.g., angular domain for mmWave
channels), users can employ sparse recovery algorithms (i.e., compressed
sensing (CS) \cite{donoho2009, Tropp2007OMP}) to recover the sparse channel
parameters and subsequently feed back the quantized version of those parameters
to the BS, e.g., \cite{Schober2019}. To design the precoding matrix, the BS
reconstructs the channels from the quantized sparse channel parameters and then
employs a conventional linear precoding scheme, e.g., maximum-ratio
transmission (MRT) or zero forcing (ZF). Such a CS-based feedback protocol in
essence adopts a separate source coding strategy of independent quantization of
each user’s channel. But, as discussed earlier, since the precoding matrix is
a function of all users' channels, information theory consideration suggests
that we can do better by adopting a DSC strategy. The main points of this paper
are to show that a deep learning framework can effectively undertake such a DSC 
design and to show by extensive numerical simulations that the proposed
learning-based precoding scheme, which bypasses explicit channel estimation, can indeed achieve a significantly better performance as compared to 
the conventional approach of separate channel estimation and
precoding, especially when downlink training and feedback resources are limited.

The second category of limited feedback precoding works is codebook based \cite{Love2008},
with discrete Fourier transform (DFT) matrix as a common choice for the
precoding codebook \cite{Nair2020}. In the training phase of the codebook-based
precoding schemes, the BS first transmits along the possible precoding
directions, and each user then sends feedback about the indices of the top-$p$
strongest received signals and their corresponding SNRs.  The feedback
information is finally processed at the BS for selecting the precoder of each
user from the codebook. If $p$ is set to be one, then the best possible 
performance of this approach is the performance of the MRT since there is no
mechanism for interference management \cite{Nair2020,Love2018}. On the other
hand, if $p$ is sufficiently large such that interference management becomes
feasible (e.g., \cite{Alkhateeb2015}), then a large amount of feedback
bits per user is required. A thorough comparison between the CSI feedback scheme and the codebook-based precoding method for MIMO systems in rich-scattering environments  
is provided in \cite{Dietl2007}. This comparison shows that CSI feedback is preferred in scenarios with very limited feedback rates. The advantage of the CSI feedback scheme in very limited downlink training and feedback resources can be even more remarkable for massive MIMO systems in limited-scattering environments. This is because CSI estimation, quantization, and feedback of a few sparse channel parameters is much more efficient in terms of downlink training and feedback resource utilization, as compared to the codebook-based precoding scheme in which the number of codewords  typically would have to scale with the number of antennas. For this reason, we only consider the comparison to the CSI feedback based precoding methods in this paper.

It is noteworthy that the use of DNNs in FDD systems with limited feedback has
been adopted in some recent works, {e.g.,
\cite{Jang2019,mashhadi2020pruning,Wen2018,Lu2019,guo2020dl,yang2020deep}}. However, these works
either focus only on a single-user scenario with no interference
\cite{Jang2019,mashhadi2020pruning,Wen2018}, or they focus on the CSI reconstruction problem at the
BS under the assumption that the perfect CSI is available at the users
\cite{Wen2018,Lu2019,guo2020dl,yang2020deep}. The work presented herein
provides a more general treatment as: i) we consider the multiuser case in which
each user can only sense and feedback its own channel, yet the precoding
process is a function of all users' channels; ii) we provide end-to-end
training of all system parameters, including downlink pilot sequences, while
accounting for CSI estimation error in order to directly enhance the downlink
spectral efficiency.

%%%%%%%%%%%%%%%%%%%%%%%%%%%%%%%%%%%%
%% I-C) Paper Organization and Notations
%%%%%%%%%%%%%%%%%%%%%%%%%%%%%%%%%%%%
\subsection{Paper Organization and Notations}
The remainder of this paper is organized as follows. Section~\ref{sec:sys} introduces the system model and the problem formulation for the end-to-end multiuser FDD downlink precoding system design. Section~\ref{sec:design} shows how to represent the FDD downlink precoding system by a neural network architecture, and further discusses how to train the neural network. Section~\ref{sec:gen} provides discussions on the generalizability of the proposed DNN, and Section~\ref{sec:gen} demonstrates the implementation details of the DNN. Section~\ref{sec:simulations} presents the simulation results. Finally, conclusions are drawn in Section~\ref{sec:conclusion}.

This paper uses lower-case letters for scalars, lower-case bold face letters for vectors and upper-case bold face letters for matrices. The real part, the imaginary part, and the dimensions of a complex matrix $\bV$ are respectively given by $\Re(\bV)$, $\Im(\bV)$, and $\dim (\bV)$. We use the superscripts $(\cdot)^T$, $(\cdot)^H$, and $(\cdot)^{-1}$ to denote the transpose, the Hermitian transpose, and the inverse of a matrix, respectively.  The identity matrix with appropriate dimensions are denoted by $\mathbf{I}$. \changeNEW{Further, $\mathbb{C}^{m\times n}$ denotes an $m$ by $n$ dimensional complex space, $\mathcal{CN}(\mathbf{0},\mathbf{R})$ represents the zero-mean circularly symmetric complex Gaussian distribution with covariance matrix $\mathbf{R}$, and $\mathcal{U}(a,b)$ represents a uniform distribution on the interval $[a,b]$.} The notations $\operatorname{Tr}(\cdot)$, $\operatorname{log}_{2}(\cdot)$, $\operatorname{log}_{10}(\cdot)$, and $\mathbb{E} [\cdot] $ represent the trace, binary logarithm, decimal logarithm, and expectation operators, respectively. Finally, $\|\cdot\|_2$ indicates the Euclidean norm of a vector.

%%%%%%%%%%%%%%%%%%%%%%%%%%%%%%%%%%%
% II) System Model
%%%%%%%%%%%%%%%%%%%%%%%%%%%%%%%%%%%
\section{System Model}
\label{sec:sys}
%%%%%%%%%%%%%%%%%%%%%%%%%%%%%%%%%%%
% II-A) Signal Model and Problem Formulation
%%%%%%%%%%%%%%%%%%%%%%%%%%%%%%%%%%%
\subsection{Signal Model and Problem Formulation}
\changeRR{Consider the downlink scenario in an FDD massive MIMO system in which a BS with $M$ transmit antennas serves $K$ single-antenna users, where $K<M$.
In this paper, we assume that the multiuser scheduling phase has already been performed at upper layer (typically based on considerations such as traffic priority, delay constraints, and user queuing status), so that in each time-frequency resource block, we have $K < M$.} Further, \editb{we} assume that the BS employs linear precoding so that the transmitted signal can be written as:
%%%%%%%%%%% 1
\begin{equation}\label{}
    \bx = \sum_{k=1}^K \bv_k s_k = \bV \bs,
\end{equation}
where $\bv_k \in \mathbb{C}^M$ is the precoding vector for the $k$-th user and forms the $k$-th column of the precoding matrix $\bV \in \mathbb{C}^{M\times K}$, which satisfies the total power constraint, i.e., $\Tr(\bV\bV^H)\leq P$, and $s_k$ is the symbol to be sent to the $k$-th user which is normalized so that $\mathbb{E}\left[\bs\bs^H\right]=\bI$. By adopting a narrowband block-fading channel model, the received signal at the $k$-th user in data transmission phase can be written as:
%%%%%%%%%%% 2
\begin{equation}\label{eq_rx_sig}
y_k =  \bh_k^H \bv_k s_k + \sum_{j\not=k} \bh_k^H \bv_ j s_j + z_k,
\end{equation}
where $\bh_k \in \mathbb{C}^M$ is the vector of downlink channel gains between the BS and user $k$ and $z_k \sim \mathcal{CN}(0,\sigma^2)$ is the additive white Gaussian noise. 
Given the received signal model at the $k$-th user in \eqref{eq_rx_sig}, the achievable rate of user $k$ is:
%%%%%%%%%%% 3
\begin{equation}\label{eq:ratek}
    R_k = \log_2\left(1 +  \frac{\lvert \bh_k^H\bv_k \rvert^2}{ \sum_{j\not=k} \lvert \bh_k^H\bv_j \rvert^2+\sigma^2 } \right).
\end{equation}

This paper aims to design the precoding matrix $\bV$ at the BS so as to maximize some network-wide utility. For simplicity, the rest of the paper uses the sum rate of the system
as the design objective, i.e.,
%%%%%%%%%%% 4
\begin{equation}\label{eq_sumrate}
R=\sum_k R_k.
\end{equation}

%%%%%%%%%%%%%%%%%%%%%%% Fig1
\begin{figure*}[t]
        \centering
        \includegraphics[width=0.95\textwidth]{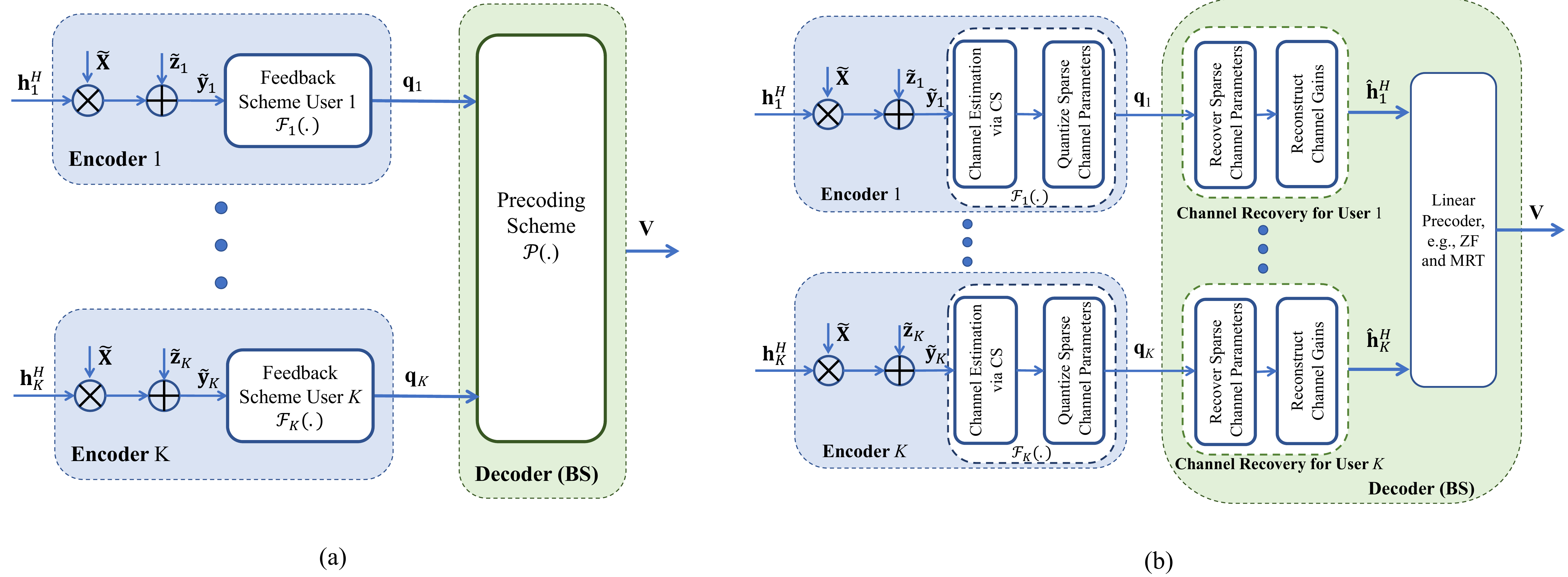}
        \caption{ {(a) The end-to-end FDD downlink precoding design problem can be viewed as a DSC problem in which the downlink training pilots and \changeNEW{the} feedback schemes adopted at the users 
        \changeNEW{can be thought of as the source encoders} and the precoding scheme adopted at the BS \changeNEW{can be thought of as the decoder.} (b) The conventional channel feedback scheme can be regarded as a separate source coding strategy of independent quantization of each user’s channel.}}
        \label{fig:Fig_Together}
\end{figure*}

In order to design the optimal precoding matrix, it is crucial for the BS to have access to instantaneous CSI. In this paper, we assume that the BS and the users have no prior knowledge of the channel state realizations, and they must acquire the CSI via downlink training and feedback. In particular, to obtain the required information for downlink precoding at the BS, we consider a downlink training phase, prior to the data transmission phase, in which the BS sends training pilots $\widetilde{\bX}\in \mathbb{C}^{M\times L}$ of length $L$, and accordingly, the $k$-th user observes $\widetilde{\by}_k\in\mathbb{C}^{1\times L}$ as:
%%%%%%%%%%% 5
\begin{equation}\label{eq_rx_pilot}
    \widetilde{\by}_k = \bh_k^H \widetilde{\bX} + \widetilde{\bz}_k,
\end{equation}
in which the transmitted pilots in the $\ell$-th pilot transmission ($\ell$-th column of $\widetilde{\bX}$) satisfies the power constraint, i.e., $\|\widetilde\bx_\ell\|_2^2\leq P$, and $\widetilde{\bz}_k \sim \mathcal{CN}(\mathbf{0},\sigma^2\mathbf{I})$ is the additive white Gaussian noise at user $k$. The $k$-th  user then seeks to obtain the useful information in \eqref{eq_rx_pilot} for the purpose of multiuser downlink precoding and subsequently to feed back this information to the BS in the form of $B$ information bits as:
%%%%%%%%%%% 6
\begin{equation}\label{eq_feedback}
\bq_k = \mathcal{F}_k\left( \widetilde{\by}_k\right), 
\end{equation}
where the function $\mathcal{F}_k: \mathbb{C}^{1\times L} \rightarrow \{\pm 1\}^B $ represents the feedback scheme adopted at user $k$.
 
Finally, the BS collects the feedback bits from all $K$ users, i.e., $\bq \triangleq [\bq_1^T,\bq_2^T,\ldots,\bq_K^T]^T$, and seeks to design the precoding matrix $\bV$ as a function of those feedback bits as:
%%%%%%%%%%% 7
   \begin{equation}\label{eq_decoding}
\bV = \mathcal{P} \left( \bq \right), 
\end{equation}
where the function $\mathcal{P}: \{\pm 1\}^{KB} \rightarrow   \mathbb{C}^{M\times K} $ represents the downlink precoding scheme. We remark that once the precoding matrix is designed at the BS, coherent detection is enabled by an additional downlink pilot transmitted along each beamforming vector to allow the receivers to calibrate detection boundaries. This procedure, which is called \textit{dedicated training phase}, typically requires only a small amount of pilot transmission \cite{Caire2010Dedicated}. In this paper, we make a simplifying assumption that the dedicated training phase is performed perfectly without any significant overhead and accordingly the rate expression in \eqref{eq:ratek} is achievable.

With the above communication models in place, the problem of maximizing the sum rate of a limited-feedback FDD system can be summarized as:
%%%%%%%%%%% 8
\begin{subequations}
\begin{align}
\label{main_problem}
\displaystyle{\Maximize_{\widetilde{\bX},\hspace{2pt}\{\mathcal{F}_k(\cdot)\}_{\forall k},\hspace{2pt}\mathcal{P}(\cdot)}} ~ & \sum_{k=1}^{K} \log_2\left(1 +  \frac{\lvert \bh_k^H\bv_k \rvert^2}{ \sum_{j\not=k} \lvert \bh_k^H\bv_j \rvert^2+\sigma^2 } \right)\\ 
\text{subject to}  \quad~ & \bV = \mathcal{P}\left(\left[\bq_1^T,\ldots, \bq_K^T \right]^T \right),\\
&\bq_k =  \mathcal{F}_k(\bh_k^H \widetilde{\bX} + \widetilde{\bz}_k), ~~\forall k, \\
& \operatorname{Tr}(\bV \bV^H) \leq P,\\
&\|\widetilde\bx_\ell\|^2_2\leq P, ~~\forall \ell,
\end{align}
\end{subequations}
in which the downlink training pilots $\widetilde{\bX}$, feedback scheme
adopted at the users $\{\mathcal{F}_k(\cdot)\}_{k=1}^{K}$, and the precoding
scheme $\mathcal{P}(\cdot)$ adopted at the BS can be optimized to enhance the
spectral efficiency. As illustrated in Fig.~\ref{fig:Fig_Together}(a), this
overall problem, which involves designing the downlink pilots, the users' 
channel estimation, quantization, and feedback schemes, jointly with the BS's 
downlink precoding scheme, can be viewed as a DSC
problem. This is because the channel estimation and quantization take place in
a distributed fashion across the users, and the feedback bits of all users are
then processed at a central node, i.e., BS, in order to construct the precoding
matrix.  This is a challenging task, because designing information
theoretically optimal DSC strategy is in general a difficult problem. Simple
heuristic, such as independent codebook-based quantization of the
channel vector at each user, is likely to be far from the optimum. The main goal
of this paper is to show that a data-driven machine learning approach can be
used to efficiently design such a DSC scheme.

%%%%%%%%%%%%%%%%%%%%%%%%%%%%%%%%%%%
% II-B) Signal Model and Problem Formulation
%%%%%%%%%%%%%%%%%%%%%%%%%%%%%%%%%%%
\subsection{Channel Model and Conventional Approaches}

This paper considers an FDD massive MIMO system operating in mmWave propagation environment\cite{pi2011introduction} in which the number of scatterers 
is limited. Accordingly, the sparse  channel of the $k$-th user is typically modeled with $L_p$ propagation paths, e.g., \cite{sohrabi2016hybrid}:
%%%%%%%%%%% 9
\begin{equation}\label{eq:channel_model}
    \bh_k = \frac{1}{\sqrt{L_p}} \sum_{\ell=1}^{L_p} \alpha_{\ell,k}\mathbf{a}_t(\theta_{\ell,k}), 
\end{equation}
where $\alpha_{\ell,k}$ is the complex gain of the $\ell$-th path between the BS and user $k$, $\theta_{\ell,k}$ is the corresponding angle of departure (AoD), and $\mathbf{a}_t\left(\cdot\right)$ is the transmit array response vector. For a uniform linear array with
$M$ antenna elements, the transmit array response vector is: 
%%%%%%%%%%% 10
\begin{equation}
\mathbf{a}_t\left(\theta \right) = \left[1,e^{j\tfrac{2\pi}{\lambda}d \sin(\theta)},\ldots,e^{j\tfrac{2\pi}{\lambda}d (M-1)\sin(\theta)}\right]^T,
\end{equation}
where $\lambda$ is the wavelength and $d$ is the antenna spacing.

The sparsity of mmWave channels in the angular domain can be exploited in
designing the feedback scheme. In particular, a conventional feedback scheme 
typically involves quantizing the estimated values of the sparse channel parameters
\cite{Schober2019}. This means that each user first employs a sparse recovery
(i.e., compressed sensing) algorithm to estimate the sparse channel parameters
then feeds back the quantized version of those parameters to the BS.
Subsequently, the BS collects the quantized channel parameters from all $K$
users, reconstructs the imperfect channel estimates based on these parameters, and
finally employs one of the conventional linear beamforming methods, e.g., MRT
or ZF, given the imperfect CSI.  Such a conventional approach typically leads
to a good performance only for systems with (i) sufficiently large pilot length
$L$ in which a decent sparse parameter estimation can be achieved via
compressed sensing, and (ii) sufficiently large number of feedback bits
$B$ where the quantization error can be made sufficiently small. 

This paper aims to show that it is possible to design \editb{an} FDD system with excellent
performance even with short training sequences and small amount of limited
feedback information bits. 
The above conventional channel feedback scheme
illustrated in Fig.~\ref{fig:Fig_Together}(b) has room for improvement, 
because it amounts to a separate source
coding strategy of independent quantization of each user's channel. However,
because the estimated channels from all the users are used jointly at the BS to
compute a downlink precoding matrix, a \emph{distributed source coding}
strategy can do better \cite{elgamal2011network,Marton1979,Nazer2007}. This is
true even if the channels to each user are uncorrelated \cite{Nazer2007}. The
design of DSC strategy is however quite challenging. In this paper, we propose
a deep learning framework to undertake such a design. In particular, we propose
a neural network architecture employing a DNN at each user to map the received
pilots directly into feedback bits, and a DNN at the BS to map the feedback
bits from all the users directly into the precoding matrix. 

\begin{figure*}[t]
        \centering
        \includegraphics[width=0.84 \textwidth]{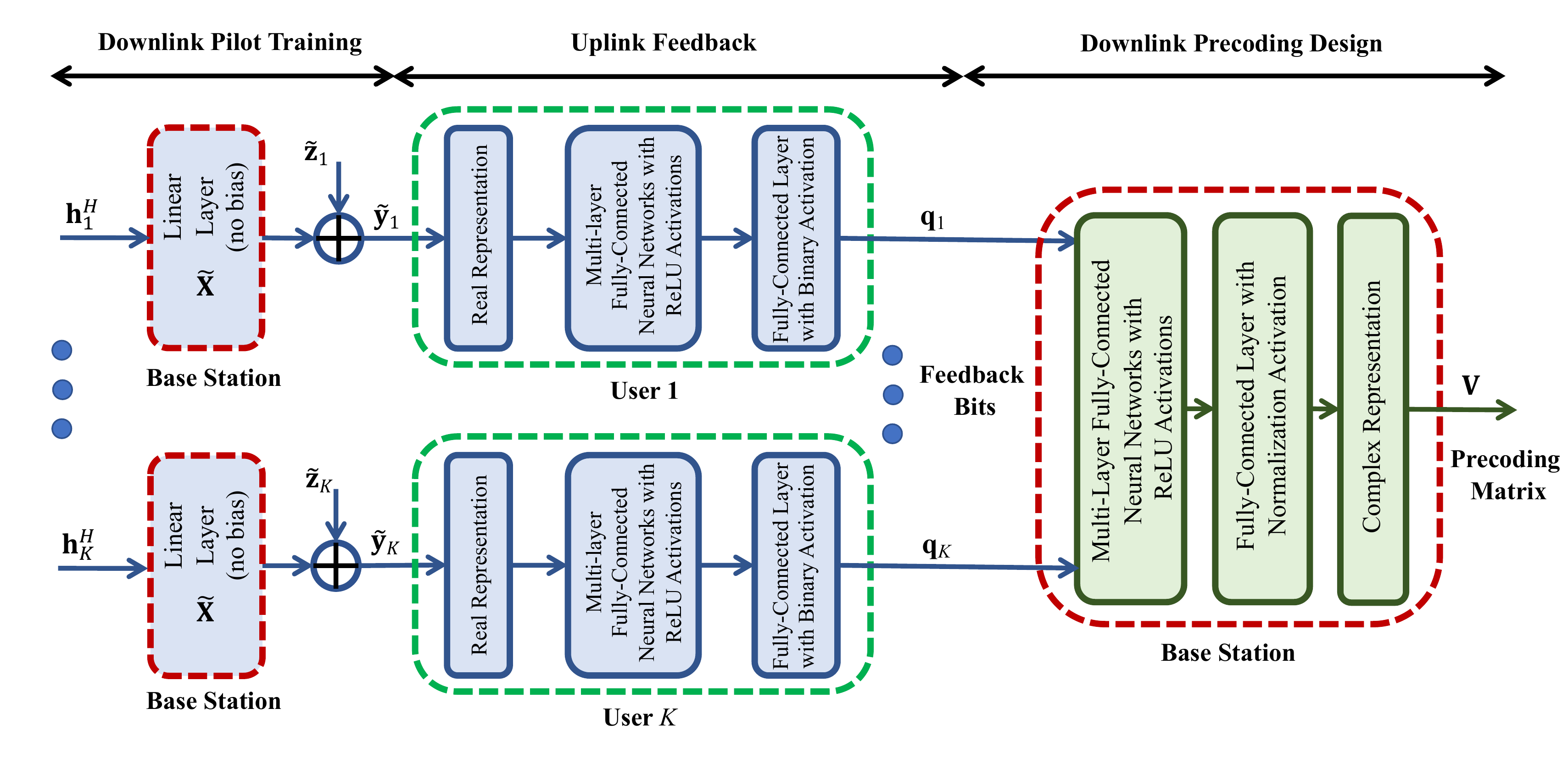}
        \caption{ {The block diagram of the proposed neural network architecture that represents an end-to-end $K$-user FDD downlink precoding system. %which involves two phases: (i) downlink training and feedback phase, and (ii) data transmission phase.
        }}
        \label{fig:DNN}
\end{figure*}

We note that the proposed neural network architecture also takes the design of downlink training pilots into account by modeling $\widetilde{\bX}$ as a linear neural network layer. The block diagram of the proposed neural network architecture that represents an end-to-end FDD downlink system is illustrated in \figurename~\ref{fig:DNN}. The details of the proposed neural network architecture are discussed in the next section.

%%%%%%%%%%%%%%%%%%%%%%% Fig2

%%%%%%%%%%%%%%%%%%%%%%%%%%%%%%%%%%%
% III) FDD Downlink Precoding System Design using Deep Learning
%%%%%%%%%%%%%%%%%%%%%%%%%%%%%%%%%%%
\section{FDD Downlink Precoding System Design using Deep Learning}
\label{sec:design}
In this section, we present how to use neural networks to model an FDD downlink system described in Section~\ref{sec:sys}, which involves two phases: (i) downlink training and uplink feedback phase, and (ii) downlink data transmission phase. Further, we discuss how to train the proposed neural network architecture in order to jointly design the downlink training pilots $\widetilde{\bX}$, the feedback scheme adopted at each user $\mathcal{F}_k(\cdot), \forall{k}$, and the downlink precoding scheme $\mathcal{P}(\cdot)$. 

%%%%%%%%%%%%%%%%%%%%%%%%%%%%%%%%%%%
% III-A) DNN Representation of DSC Strategy
%%%%%%%%%%%%%%%%%%%%%%%%%%%%%%%%%%%
\subsection{DNN Representation of DSC Strategy}

In order to represent the FDD downlink precoding system described in Section~\ref{sec:sys} as a neural network, we need to model downlink pilot transmission and the users' operations in the downlink training phase as well as the BS's operations in the data transmission phase. 

%%%%%%%%%%%%%%%%%%%%%%%%%%%%%%%%%%%
% Downlink Pilot Training
%%%%%%%%%%%%%%%%%%%%%%%%%%%%%%%%%%%
\subsubsection{Downlink Pilot Training}
We begin by modeling the first part of the downlink training phase, i.e., downlink pilot transmission, in which the BS sends training pilots $\widetilde{\bX}\in \mathbb{C}^{M\times L}$ in $L$ downlink transmissions, and consequently the $k$-th user observes $\widetilde{\by}_k = \bh_k^H \widetilde{\bX} + \widetilde{\bz}_k$. By considering $\bh_k^H$ as the input, it is easy to see that the received signal at each user in the downlink training phase can be modeled as the output of a fully-connected neural network layer with linear activation function, in which the weight matrix is $\widetilde{\bX}$ and the bias vector is zero, followed by an additive zero-mean noise with variance $\sigma^2$.

To ensure that the designed weight matrix $\widetilde{\bX}$ satisfies the per-transmission power constraint $P$, we adopt a weight constraint under which each column of $\widetilde{\bX}$ satisfies $\|\widetilde\bx_\ell\|^2_2 \le P$. 
We remark that in the machine learning literature weight constraints are employed as means of regularization to reduce overfitting in DNNs, e.g., \cite{hinton2012improving}. However, in this paper, we adopt a particular choice of weight constraint explained above to model the physical constraint on the transmit power level of a cellular BS. \changeRR{The further implementation details are provided in Section~\ref{sec:implementation}.}  

%%%%%%%%%%%%%%%%%%%%%%%%%%%%%%%%%%%
% Uplink Feedback
%%%%%%%%%%%%%%%%%%%%%%%%%%%%%%%%%%%
\subsubsection{Uplink Feedback}
Upon receiving $\widetilde{\by}_k$ in the downlink training phase, the main objective of user $k$ is to summarize its observation from $\widetilde{\by}_k$ and to feed back that summary as $B$ information bits to the BS  for the purpose of downlink precoding. This procedure can be represented by an $R$-layer fully-connected DNN in which the feedback bits of user $k$ can be written as:
%%%%%%%%%%%%%%% 11
\begin{align}\label{eq_NN1}  \nonumber
\bq_k &=\\
&\operatorname{sgn}\left(\bW^{(k)}_{R}\sigma_{{R-1}}\left( \cdots  \sigma_{1}\left( \bW_{1}^{(k)} \bar{\by}_k + \bb_{1}^{(k)} \right)  \cdots \right)+\bb_{R}^{(k)} \right), 
\end{align}
where $\bq_k \in \{ \pm 1 \}^B$,
$\{\bW_{r}^{(k)},\bb_{r}^{(k)} \}_{{r} = {1}}^{
R}$ is the set of the trainable parameters for user k,
$\sigma_{r}$ is the activation function for the $r$-th layer, and the sign function $\operatorname{sgn}(\cdot)$ is the activation function of the last layer to generate bipolar feedback bits for each component of $\bq_k$. \changeRR{In this paper, we adopt the rectified linear unit (ReLU) activation function at the hidden layers, i.e., $\sigma_r(\cdot) = \max(\cdot, 0)$.} In \eqref{eq_NN1}, the real representation of $\widetilde{\by}_k$, i.e.,
%%%%%%%%%%%%%%% 12
\begin{equation}
  \bar{\by}_k \triangleq \left[\Re\left( \widetilde{\by}_k\right) , \Im\left( \widetilde{\by}_k\right) \right]^T,    
\end{equation}
 is considered as the input of the DNN since most of the existing deep learning libraries only support real-value operations.
Further, the dimensions of the trainable weight matrices and the bias vectors in \eqref{eq_NN1} are respectively:
%%%%%%%%%%%%%%% 13
\begin{align}
\dim{(\bW_{r})} =   
\begin{cases} 
 \ell_{r}\times 2L,
& {r} = {1},\\ 
\ell_{r}\times\ell_{r-1},
& {r} = {2},\ldots,{R}-{1},\\ 
B\times\ell_{r-1}, & {r}={R},\end{cases}
\end{align}
and
%%%%%%%%%%%%%%% 14
\begin{align}
\dim{(\bb_{r})} =   
\begin{cases} 
 \ell_{r}\times1,
& r = 1,\ldots,R-1,\\ 
B\times1, & r=R,\end{cases}
\end{align}
where $\ell_{r}$ is the number of neurons in the $r$-th hidden layer.

%%%%%%%%%%%%%%%%%%%%%%%%%%%%%%%%%%%
% Downlink Precoding Design
%%%%%%%%%%%%%%%%%%%%%%%%%%%%%%%%%%%
\subsubsection{Downlink Precoding Design}
Under the assumption of an error-free feedback channel between each user and the BS, the BS collects the information bits of all users, then designs the precoding vectors as a function of these information bits. Analogous to the user side, the operations at the BS can be modeled by another DNN with $T$ dense layers, where the $t$-th layer includes $\ell_t^{\prime}$ neurons. In particular, 
the real representation of the collection of the precoding vectors, i.e.,
%%%%%%%%%%%%%%% 15
\begin{equation} 
\bv=\left[\operatorname{vec}\left(\Re\left({\bV}\right)\right)^T,\operatorname{vec}\left(\Im\left({\bV}\right)\right)^T\right]^T,
\end{equation}
can be written as:
%%%%%%%%%%%%%%% 16
\begin{equation}
\bv = {\widetilde{\sigma}}_{T}\left({\widetilde{\bW}}_{T}{\widetilde{\sigma}}_{{T-1}}\left( \cdots \widetilde\sigma_{1}\left( \widetilde\bW_{1} \bq + \widetilde\bb_{1} \right) + \cdots \right)+\widetilde\bb_{T} \right),
\end{equation}
where $\widetilde\sigma_{t}$, $\widetilde\bW_{t}$, and $\widetilde\bb_{t}$ are the activation function, the weights, and the biases in the $t$-th layer, respectively, and the collection of feedback bits of all $K$ user, i.e., $\bq = [\bq_1^T,\ldots,\bq^T_K]^T$, is the input vector to the DNN. In order to ensure that the total power constraint is satisfied, a normalization layer with activation function:
%%%%%%%%%%%%%%% 17
\begin{equation}
\changeNEW{{\widetilde\sigma_{T}(\cdot)= \sqrt{P} \tfrac{\cdot}{\|\cdot\|_2}}}
\end{equation}
is employed at the last layer of the DNN. \changeRR{For the other layers of the BS's DNN, we adopt the ReLU activation function.}

\changeRR{We remark that if some prior knowledge about the channel, such as sparsity level of the channel, angular domain information, and pathloss, is available, we can consider them as the input to the user-side DNNs and/or BS-side DNN, and possibly enhance the performance of the overall network. However, in this paper, we consider the most challenging setup where the BS has no explicit information about the channel parameters, and it should learn the statistics of those parameters from the training data.}

The block diagram of the overall proposed neural network architecture that represents an end-to-end two-phase FDD downlink precoding system is illustrated in Fig.~\ref{fig:DNN}. In this neural network the trainable parameters are the training pilot matrix $\widetilde{\bX}$, the DNN parameters  $\Theta_\text{R}^{(k)} \triangleq\{\bW_{r}^{(k)},\bb_{r}^{(k)} \}_{{r} = {1}}^{
R}$ at the user side, and the DNN parameters  $\Theta_\text{T} \triangleq \{\bW_{t},\bb_{t}\}_{{t} = {1}}^{T}$ at the BS side. 

%%%%%%%%%%%%%%%%%%%%%%%%%
%   III-B) DNN Training with a Hidden Binary Layer
%%%%%%%%%%%%%%%%%%%%%%%%%
\subsection{DNN Training with a Hidden Binary Layer}

%By representing a limited-feedback FDD downlink precoding system with the 
We now describe the training of the DNN architecture in Fig.~\ref{fig:DNN} for
the sum rate maximization objective as stated in the following:
%%%%%%%%%%%%%%% 18
\begin{equation}\label{eq_training_prob}
\max_{\widetilde{\bX}, \{\Theta_\text{R}^{(k)}\}_{k=1}^K,\Theta_\text{T} } \hspace{-6pt}\mathbb{E}_{\bH,\widetilde{\bz}}\left[ \sum_k 
\log_2\left(1 +  \frac{\lvert \bh_k^H\bv_k \rvert^2}{ \displaystyle{\sum_{j\not=k}} \lvert \bh_k^H\bv_j \rvert^2+\sigma^2 } \right)
\right],
 \end{equation} 
where the expectation is over the distribution of the channels, i.e., $\bH \triangleq [\bh_1,\ldots,\bh_K]^H$, and the distribution of the noise in the downlink training phase, i.e., $\widetilde{\bz}\triangleq [\widetilde{\bz}_1,\ldots,\widetilde{\bz}_K]^T$. 
The parameter space consists of the training pilot matrix, the users' feedback schemes, and the BS's precoding scheme.  

We assume certain distributions of the channels and the noise in the downlink training phase and accordingly generate a large set of channel and noise realizations for the training purpose. The training problem for \eqref{eq_training_prob} can
then be efficiently tackled by employing stochastic gradient descent (SGD)
algorithms in which the expectation in \eqref{eq_training_prob} is approximated
with the empirical average over the training samples. SGD-based training algorithms
require partial derivatives of the loss function, here the sum rate expression,
with respect to all the trainable parameters in order to update the parameters
in each iteration. The partial derivatives are computed via
\textit{back-propagation} method which is an efficient implementation of the
chain rule in directed computation graphs.

Due to the fact that the derivative of the output of a binary thresholding
neuron is zero almost everywhere (with the exception of the origin where the 
function is not differentiable), the conventional back-propagation method
cannot be directly used to train the neural layers prior to that binary layer.
A common practice in the machine learning literature to overcome this issue is to
approximate the activation function of a binary thresholding layer by another
smooth differentiable function during the back-propagation phase
\cite{hinton2012videos, bengio2013, chung2016}. Such approximation of a binary
layer in the back-propagation phase, which is \changeNEW{first introduced in
\cite{hinton2012videos},} is known as straight-through (ST) estimation.
A variant of the ST estimator, called sigmoid-adjusted ST, is to replace
the derivative factor with the gradient of the function $2
\operatorname{sigm}(u) -1$, where $\operatorname{sigm}(u) = 1/({1+\exp(-u)})$
is the sigmoid function.  It is shown in \cite{chung2016} that the performance
of the sigmoid-adjusted ST estimator can be further improved by adopting the
slope-annealing trick, in which the slope of the sigmoid function is gradually
increased as training progresses. In particular, the sigmoid-adjusted ST with
slope annealing estimator approximates the sign function
$\operatorname{sgn}(u)$ in the back-propagation phase with a properly scaled
sigmoid function as:
%%%%%%%%%%%%%%%%%%%%%%% 19
\begin{equation}
2 \operatorname{sigm}(\alpha^{(i)}u) -1 = \frac{2}{1+\exp(-\alpha^{(i)}u)} -1,
 \end{equation}
 where $\alpha^{(i)}$ is the annealing factor in the $i$-th epoch satisfying $\alpha^{(i)} \geq \alpha^{(i-1)}$. 
In this paper, we adopt the sigmoid-adjusted ST with annealing during the back-propagation phase to compute the gradients of the binary layer considered at the last  stage of the user side in  the neural network in Fig.~\ref{fig:DNN}.
Further implementation details are provided in Section~\ref{sec:implementation}.

%%%%%%%%%%%%%%%%%%%%%%%%%
%   IV) Generalizability of the Proposed DNN
%%%%%%%%%%%%%%%%%%%%%%%%% 
\section{Generalizability}
\label{sec:gen}

%%%%%%%%%%%%%%%%%%%%%%%%%
%   IV.A) Towards Robust DNN Design
%%%%%%%%%%%%%%%%%%%%%%%%% 
\subsection{Towards Robust DNN Design}

The training of the proposed DNN architecture assumes a specific channel environment.
The natural question that arises for any deep learning-based algorithm,
is then how generalizable the proposed DNN is, if it is trained under one
set of system parameters, but tested under a different set of parameters. 
For the problem under consideration, we can group the system parameters into two
categories. The first category consists of the parameters that only change the
input distribution, e.g., the channel parameters and the noise statistics. For
these parameters, training under a variety of system parameters typically
enhances robustness. By considering the number of channel paths $L_p$ as an
example, we numerically show in Section~\ref{sec:simulations} that by training
the proposed DNN over different number of paths, the robustness of the proposed
DNN can be enhanced. However, for the second category of the system parameters,
which also change the input/output dimensions of some layers of the DNN,
training the DNN to operate for different system dimensions is more
challenging. In this case, modification of the proposed DNN architecture and/or
devising novel training procedures are needed. In the next two subsections, we 
explain how to enhance the generalizability of the proposed DNN with 
respect to the number of feedback bits $B$ and the number of users $K$.      

%%%%%%%%%%%%%%%%%%%%%%%%%
%   IV.B) Towards Generalizability for $B$
%%%%%%%%%%%%%%%%%%%%%%%%% 
\subsection{Towards Generalizability for $B$}
\label{sec:gen_B}

In the proposed neural network architecture in Fig.~\ref{fig:DNN}, it appears that we have to
train a different DNN for each value of feedback rate limit $B$ since the
dimension of the output of each user's DNN and accordingly the dimension of the
input to the BS's DNN are determined by the value of $B$. This would
be a tedious task for practical implementation if the amount of feedback can
possibly vary. In practical system design, it is desirable to train
a common neural network that can operate over a wide range
of feedback capacities. To address this need, we propose the following two-step
training approach. We propose to first train a modified version of the proposed
neural network in which the output of the user-side DNN is not binary anymore and
instead is soft binary valued, i.e., real numbers belongs to $[-1,1]$,
generated by $S$ neurons with hyperbolic tangent ($\operatorname{tanh}$)
activation functions. After this modified network has been trained, we obtain
the empirical probability distribution function (PDF) of the output of the $\operatorname{tanh}$
layer, then design the optimal scalar quantizer for this distribution for
different values of quantization bits (i.e., $Q$) according to the Lloyd-Max
algorithm. After the encoder parameters including the training pilot
sequences and the user-side DNN parameters are obtained, as the second step of the
training procedure, we seek to design the decoder parameters at the BS to
generate the precoding matrix. In particular, the BS receives
a $Q$-bit quantized version of $S$ soft binary signals from each user, and
the task of the BS-side DNN is to map these $K\times S$ quantized signals to the
precoding matrix such that the average sum rate is maximized. The weights and
biases of the BS-side DNN can be learned using the SGD-based training. Note that
in this scheme the amount of feedback per user is equal to
$B=S\times Q$, hence by varying different quantization levels $Q$,
we can train the same DNN to operate for 
different values of $B$.  

%%%%%%%%%%%%%%%%%%%%%%%%%
%   IV.C) Towards Generalizability for $K$
%%%%%%%%%%%%%%%%%%%%%%%%% 
\subsection{Towards Generalizability for $K$}
\label{sec:gen_K}

The proposed architecture has a separate DNN at each user. At a first glance, 
it may appear that they need to be trained separately; further, their trained 
parameters would also depend on the total number of users in the system. 
\editb{However, our experimental results suggest that
in some scenarios where the channel
distribution for different users are i.i.d.}, one common user-side DNN can be
adopted, regardless the number of users in the system.  Motivated by this
observation, we propose to first learn the encoding parameters (including the
pilot matrix and the channel estimation/feedback scheme) by training a single-user system. Then we adopt the same encoding DNN for all users, and only train the parameters of
the BS-side DNN separately depending on the total number of users in the system.
Such a design is much more efficient than training different user-side DNNs at
different users for different systems.  Only at the BS side, we need to train 
and store different DNNs for handling different numbers of users in the network.

%%%%%%%%%%%%%%%%%%%%%%%%%
%   V) Implementation Details
%%%%%%%%%%%%%%%%%%%%%%%%%
\section{Implementation Details}
\label{sec:implementation}

We implement the proposed neural network in Fig.~\ref{fig:DNN} using two open-source deep learning libraries, namely TensorFlow \cite{tensorflow2016} and Keras \cite{chollet2015}. \changeRR{We then follow the training procedure in Algorithm~1 to learn the parameters of the proposed DNN. In this section, we provide the implementation details of the proposed DNN in Fig.~\ref{fig:DNN} and its training procedure in Algorithm~1.} 

We employ a variant of the SGD-based training method, called \textit{Adam optimizer} \cite{adam2014}, 
with a mini-batch size of \changeRR{$N_\text{b} = 1024$} and a learning rate \changeRR{$\eta$}
progressively decreasing from $10^{-3}$ to $10^{-5}$.
Unless otherwise stated, we use $4$-layer fully-connected DNN at the user side as well at the BS side, i.e., $R=T=4$, 
while the number of hidden neurons of different layers at the user side and at the BS side 
 are $[\ell_1,\ell_2,\ell_3,\ell_4]=[1024,512,256,B]$ and $[\ell_1^\prime,\ell_2^\prime,\ell_3^\prime,\ell_4^\prime]=[1024,512,512,2MK]$, respectively. 
For faster convergence, each dense layer is preceded by a batch normalization layer \cite{Ioffe2015}. In the simulations, we slowly increase the annealing parameter 
in the $i$-{th} epoch according to $ \alpha^{(i)} = \max\{1.001 \alpha^{(i-1)},10\}$ where $\alpha^{(0)}=0.5$ and each epoch consists of \changeRR{$N_\text{batch}=200$} mini-batches. 

%%%%%%%% Algorith 1
\begin{algorithm}\label{alg:DNN}
\changeRR{
\SetAlgoLined
 \underline{\textbf{Input:}} $K, L, M, B, \sigma^2$ // System parameters\;
 \hspace{26pt} $N_{\text{v}}, N_{\text{b}}$ \hspace{12pt}// Validation set size and batch size\;
 \hspace{26pt} $N_\text{batch}$ \hspace{16.5pt}// Number of batches per epoch\;
 \hspace{26pt} $N_\text{ep,max}$ \hspace{12pt}// Max epochs with no improvement\;
 \hspace{26pt} $\alpha^{(0)}$, $\eta^{(0)}$ // Initial annealing and learning rates\;
 
 \underline{\textbf{Initialization:}} \\
  \hspace{20pt} $i \leftarrow 0$\; \hspace{20pt} \textbf{Generate} validation set $\mathcal{S}_{\text{v}}$ of size $N_{\text{v}}$\;
  \hspace{20pt} \text{best\_rate} $\leftarrow$ average sum rate on $\mathcal{S}_{\text{v}}$\;
  \underline{\textbf{DNN Training:}}\\
  \While{$i < N_{\text{ep,max}}$}{
 \For{$ t = 1 \ldots, N_\text{batch}$}{
  \textbf{Generate} a training mini-batch $\mathcal{S}^{(t)}_b$ of size $N_b$\;
  \textbf{Update} weights and biases using GD (with the annealed sigmoid-adjusted ST approximation for the binary layer) on $\mathcal{S}^{(t)}_b$\;}
  \text{current\_rate} $\leftarrow$ average sum rate on $\mathcal{S}_{\text{v}}$\;
  
  \uIf{\text{current\_rate} $>$ \text{best\_rate}}{\textbf{Save} $\{ \mathbf{W}_r^{(k)}, \mathbf{b}_r^{(k)}\}_{r = 1}^R,\hspace{2pt}\forall k$, $\{ \tilde{\mathbf{W}}_t, \tilde{\mathbf{b}}_t\}_{t = 1}^T$, $\tilde{\mathbf{X}}$\;
\text{best\_rate} $\leftarrow$ \text{current\_rate}\;
  $i \leftarrow 0$, ~~~~// reset no-improvement counter\;
  }\Else{$i \leftarrow i + 1$, // update no-improvement counter\;} 
  \textbf{Increase} annealing rate\;
  \textbf{Decrease} learning rate\;
 }
 \caption{Training procedure of the proposed DNN to design an FDD downlik precoding system}}
\end{algorithm}

\changeRR{In order to optimize the downlink training pilot matrix $\widetilde{\bX}$, we define $\widetilde{\bX}$ as a training variable in TensorFlow whose initial value is randomly generated according to i.i.d.\ complex Gaussian distribution with zero mean and variance $\sqrt{P/M}$ such that the transmitted pilots in the $\ell$-th pilot transmission satisfy the power constraint, i.e., $\|\widetilde\bx_\ell\|^2_2\leq P$. To ensure that the final designed pilot matrix also satisfies such a power constraint, we always normalize the updated $\widetilde{\bX}$ in each iteration such that $\|\widetilde\bx_\ell\|^2_2 =  P$.}

We fix the distribution of the channels as well as
the distribution of the noise in the downlink training phase, so that 
we can generate as many data samples as needed for training the DNN. This
assumption enables us to investigate the ultimate performance of the deep
learning-based precoding design for FDD systems with limited feedback. The
investigation on the minimal size of the training data set that leads to a reasonable
performance is a direction for future work.

We monitor the generalization performance of the DNN during training by computing the achieved average sum rate by the DNN for a holdout validation data set of \changeRR{$N_\text{v} = 10^4$} samples, and keep the model parameters that have achieved the best generalization performance so far. The training procedure is terminated when the generalization performance for the validation data set has not improved over a large number of (e.g., \changeRR{$N_\text{ep,max}=300$}) epochs. After the proposed DNN in Fig.~\ref{fig:DNN} is trained, we have access to the design of the training pilot matrix $\widetilde{\bX}$, the feedback scheme for each user $\mathcal{F}_k(\cdot)$, and the BS's precoding scheme $\mathcal{P}({\cdot})$.

%%%%%%%%%%%%%%%%%%%%%%%%%%%%%%%%%%%%%%
% VI) Numerical Results
%%%%%%%%%%%%%%%%%%%%%%%%%%%%%%%%%%%%%%
\section{Numerical Results}
\label{sec:simulations}
We now illustrate the performance of the proposed deep learning-based precoding
method for FDD systems with limited feedback. We compare the performance of the
proposed deep learning-based precoding framework with that of the conventional
MRT and ZF precoding methods. For each of the MRT and ZF precoding baseline
methods we consider four different system settings: (i) perfect CSI at the BS
and the users; (ii) perfect CSI at the users with limited-feedback links
between each user and the BS; (iii) no prior CSI available at either the BS or
the users with infinite-capacity feedback links; and (iv) no prior CSI
available at either the BS or the users with limited-capacity feedback links.
Before presenting the numerical results, we first provide a brief explanation
of each of the baseline methods.

%%%%%%%%%%%%%%%%%%%%%%% Fig4 ---> Fig3
\begin{figure*}[t]
        \centering
        \includegraphics[width=0.9 \textwidth]{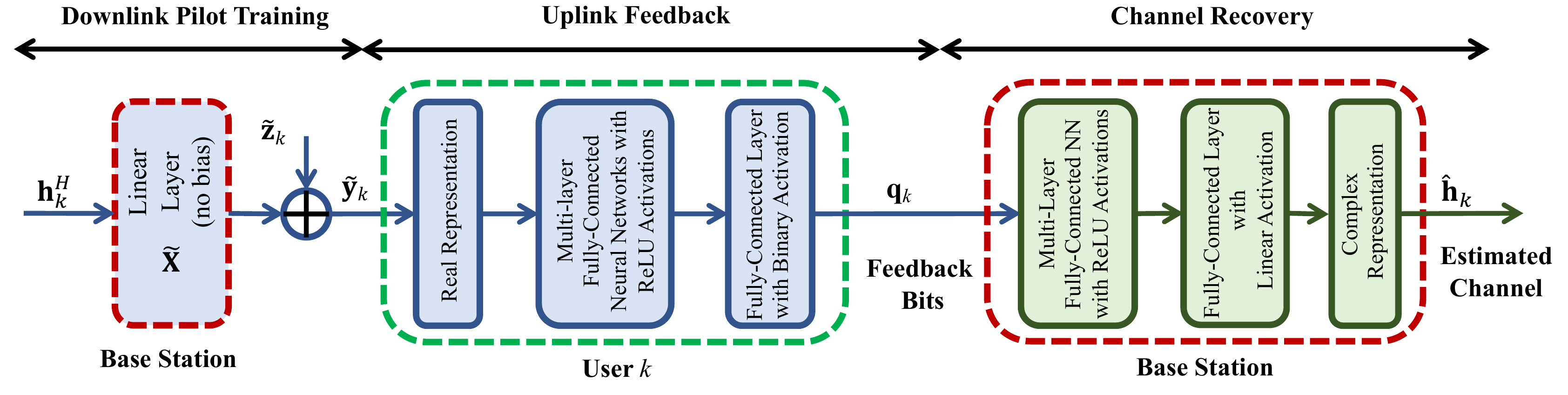}
        \caption{ {The block diagram of the neural network architecture which models the idea of per-user channel estimation. The estimated channels obtained by the trained DNN are then used to design the linear precoders.}}
        \label{fig:DNN_separate}
\end{figure*}

%%%%%%%%%%%%%%%%%%%%%%%%%%%%%%%%%%%%%%
% VI-A) Baseline Methods
%%%%%%%%%%%%%%%%%%%%%%%%%%%%%%%%%%%%%%
\subsection{Baseline Methods}
%%%%%%%%%%%%%%%%%%%%%%%%%%%%%%%%%%%%%%
% MRT/ZF with CSIT:
%%%%%%%%%%%%%%%%%%%%%%%%%%%%%%%%%%%%%%
\subsubsection{MRT/ZF with Full CSI at Transmitter (CSIT)} If full CSI is available at the transmitter (e.g., the BS), conventional multiuser linear precoding schemes such MRT or ZF can be used, in which the precoding matrix is respectively given by:
%%%%%%%%%%%%%%%%%%%% 20,21
\begin{eqnarray} \label{eq:MRT}
\bV_\text{MRT} &=& \gamma_\text{MRT} \bH^H,\\
\bV_\text{ZF} &=& \gamma_\text{ZF} \bH^H (\bH \bH^H)^{-1} ,
\label{eq:ZF}
\end{eqnarray}
where $\bH$ is the instantaneous CSI, and $\gamma_\text{MRT}$ and $\gamma_\text{ZF}$ are constants ensuring that the power constraint is satisfied.

%%%%%%%%%%%%%%%%%%%%%%%%%%%%%%%%%%%%%%
% MRT/ZF with CSIR and finite-capacity feedback:
%%%%%%%%%%%%%%%%%%%%%%%%%%%%%%%%%%%%%%
\subsubsection{MRT/ZF with Full CSI at Receiver (CSIR) and Finite-Capacity Feedback}
Under the perfect CSI assumption at the user side, user $k$ has a perfect knowledge about its sparse channel parameters, i.e., $\{\alpha_{\ell,k},\theta_{\ell,k} \}_{\forall \ell=1}^L$. User $k$ aims to transfer this knowledge to the BS by sending the index of the quantized version of the channel parameters, $\{\hat{\alpha}_{\ell,k},\hat{\theta}_{\ell,k} \}_{\forall \ell=1}^L$, over an error-free $B$-bit finite-capacity feedback link. By using the channel model in \eqref{eq:channel_model},  the BS can then reconstruct the estimated channel vectors, $\hat{\bh}_k, \forall k$, from those quantized channel parameters. 
Finally, the MRT and ZF precoder matrices can be computed respectively via expressions \eqref{eq:MRT} and \eqref{eq:ZF} in which the instantaneous CSI matrix $\bH$ is replaced with the matrix of channel estimates 
$\hat{\bH} \triangleq [\hat{\bh}_1,\ldots,\hat{\bh}_K]^H$. In the simulations, we assume that each user allocates $\tfrac{B}{3L_p}$ quantization bits to each real parameter of the channel, i.e, $\{\Re({\alpha_{\ell,k}}),\Im({\alpha_{\ell,k}}),\theta_{\ell,k} \}_{\forall \ell=1}^{L_p}$. Further, by assuming that the distribution of the channel parameters is known at the BS and the user, we consider the optimal quantization scheme for the sparse channel parameters obtained by the Lloyd-Max quantization algorithm.

%%%%%%%%%%%%%%%%%%%%%%%%%%%%%%%%%%%%%%
% MRT/ZF with DL training and infinite-capacity feedback:
%%%%%%%%%%%%%%%%%%%%%%%%%%%%%%%%%%%%%%
\subsubsection{MRT/ZF with Downlink (DL) Training and Infinite-Capacity Feedback}
In this baseline, no prior channel knowledge is initially assumed at either BS or the users. However, each user estimates its corresponding channel vector in the DL training phase and subsequently feeds the estimated channel back to the BS over an infinite-capacity link. As a result, the BS constructs the MRT/ZF precoder based on the estimated channels at the users in the downlink training phase. For estimating sparse channels in mmWave environment, user $k$ seeks to recover the sparse channel parameters by employing CS techniques. Here, we adopt a widely-used CS algorithm orthogonal matching pursuit (OMP) \cite{Tropp2007OMP}. 

%%%%%%%%%%%%%%%%%%%%%%%%%%%%%%%%%%%%%%
% MRT/ZF with DL training and finite-capacity feedback:
%%%%%%%%%%%%%%%%%%%%%%%%%%%%%%%%%%%%%%
\subsubsection{MRT/ZF with DL Training and Finite-Capacity Feedback}

Similar to the previous case, the channel parameters are first estimated at each user independently, then quantized versions of the parameters are fed back to the BS via finite-capacity feedback links. Different DL training methods are investigated, e.g., OMP for sparse channel recovery. But in particular, to illustrate the gain of the proposed DSC strategy as compared to the separate source coding strategy, we implement a scheme that uses a deep learning strategy to perform channel estimation, quantization, feedback and reconstruction at the BS,   followed by conventional linear precoding. We train a neural network architecture as shown in Fig.~\ref{fig:DNN_separate} to jointly design the training pilots, feedback scheme of the user, and the channel reconstruction function at the BS so that the reconstructed channels at the BS for all users are as accurate as possible under the finite capacity feedback constraint. Accordingly, we choose the average mean squared error (MSE) of the channels as the loss function, i.e., $\mathcal{L}=\mathbb{E}\left[\| \hat{\bh}_k - \bh_k\|_2^2\right]$. For implementing this neural network, we follow the same implementation strategies described in Section~\ref{sec:implementation}. 
The estimated channels are then used to compute the MRT and ZF precoders.
Note that the channels for different users are i.i.d., so the same DNN structure can be used for all users.

%%%%%%%%%%%%%%%%%%%%%%%%%%%%%%%%%%%%%%
% VI-B) Performance Comparison to Baseline Methods
%%%%%%%%%%%%%%%%%%%%%%%%%%%%%%%%%%%%%%
\subsection{Performance Comparison to Baseline Methods}
\label{sec:proof_of_concept}
The numerical simulations in this subsection demonstrate the performance of the proposed deep learning-based precoding for an FDD system as compared to the described baseline methods. Here, we consider an FDD massive MIMO system in which a BS with $M=64$ half-wavelength-spaced antennas serves $K=2$ users in an $L_p=2$ path environment. We assume that the complex gain of each path is modeled by a Gaussian distributed random variable, i.e., $\alpha_{\ell,k} \sim \mathcal{CN}(0,1)$, and the corresponding AoD is modeled by a uniform distributed random variable, i.e., $\theta_{\ell,k} \sim \mathcal{U}(-30^\circ,+30^\circ)$. We set the signal-to-noise-ratio as $\operatorname{SNR} =10\log_{10}(\tfrac{P}{\sigma^2}) = 10$dB. We note that some system parameters are chosen to be relatively small, e.g., the number of users and the number of channel paths, such that we would be able to train the proposed neural network under different settings in a relatively short period of time. While the numerical results for such relatively small system parameters can indeed serve as a proof of concept for the proposed deep learning-based precoding approach, we also provide discussions and simulations on how to make the proposed architecture more generalizable to larger system parameters.

\begin{figure}[t]
        \centering
        \includegraphics[width=0.5\textwidth]{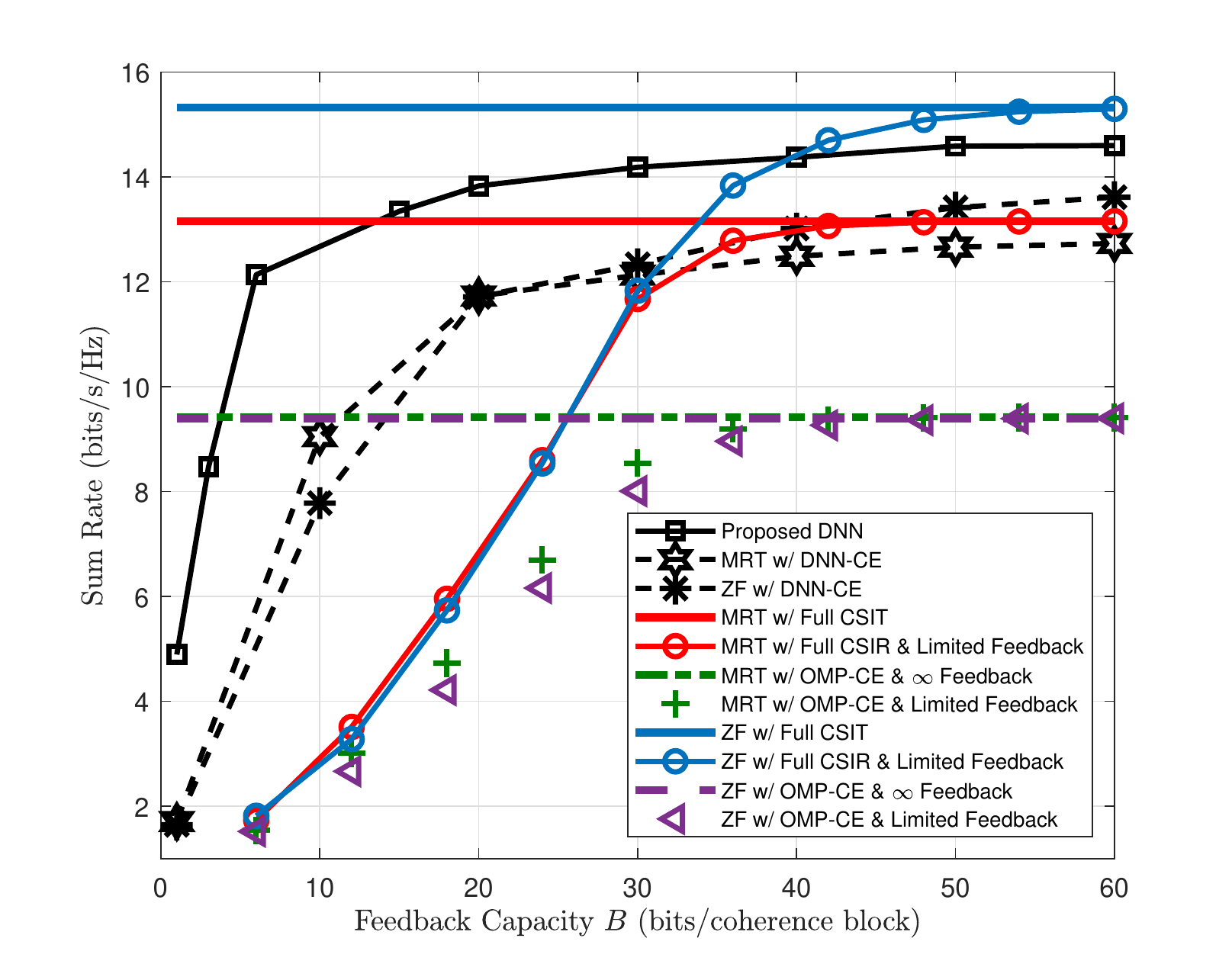}
        \caption{Sum rate achieved by different methods in a 2-user FDD system with $M=64$ and $L =8$.}
       \label{fig:L8}
\end{figure}

As the first experiment, we plot the average sum rate\footnote{In all numerical simulations, we report the average sum rate performance of different methods over $10^4$ channel realizations.} against per-user feedback capacity $B$ for a system in which the pilot length $L=8$ is much smaller than the number of antennas $M=64$. Fig.~\ref{fig:L8} shows that the proposed deep learning-based precoding approach with only $15$-bit finite-capacity feedback link already outperforms the MRT precoding even for the scenario where MRT precoder is designed based on the perfect CSI. Note that the MRT precoding approach only seeks to maximize the useful signal power without having a mechanism to mange the interference. So, achieving a better performance by the proposed approach as compared to MRT suggests that the trained DNN has learned a mechanism to reduce the inter-user interference in a multiuser FDD system.

Further, Fig.~\ref{fig:L8} illustrates that the proposed method significantly outperforms the MRT/ZF precoding schemes with downlink training and conventional OMP channel sparse recovery. This implicitly means that for scenarios that the number of channel observations is very limited (e.g., $L=8$) and hence, a good sparse recovery may not be feasible, the optimal feedback scheme is quite different from first estimating then quantizing the sparse channel parameters. The numerical results in Fig.~\ref{fig:L8} indicates that the proposed deep learning-based precoding method is indeed an efficient framework to design a better feedback scheme for such scenarios.

%%%%%%%%%%%%%%%%%%%%%%% Fig5 ---> Fig4

In Fig.~\ref{fig:L8}, we also illustrate the performance of a deep learning-based approach that implements the idea of channel recovery followed by linear MRT/ZF precoding. Fig.~\ref{fig:L8} shows that in short-pilot-length scenarios such as $L=8$, a DNN with channel recovery-based approach can outperform the \changeNEW{conventional} CS-based recovery methods. This is another confirmation showing that the approach of first estimating then quantizing the sparse channel parameters is quite suboptimal for limited-$L$ scenarios. 

Moreover, Fig.~\ref{fig:L8} shows that the proposed DNN which directly optimizes the sum rate and bypasses the channel recovery procedure achieves a higher rate as compared to the DNN that first recovers each user's channels and then applies MRT/ZF 
at the same feedback capacity constraint. To quantify the gain of the proposed DSC strategy as compared to an optimized separate source coding strategy, we note that if operating at $12$ bits/s/Hz sum rate (about $80\%$ of the sum rate achieved with full CSIT), the DSC strategy requires only about 7-bits of feedback, as compared to 
more than $20$-bits of feedback for the separate source coding strategy, whereas both are designed using DNN.
Fig.~\ref{fig:L8} also shows that the proposed precoding method with about $20$-bit finite-capacity feedback links can already achieve almost $90\%$ of the sum rate in the ZF precoding method with perfect CSIT. 

%%%%%%%%%%%%%%%%%%%%%%% Fig6 ---> Fig5
\begin{figure}[t]
        \centering
        \includegraphics[width=0.5\textwidth]{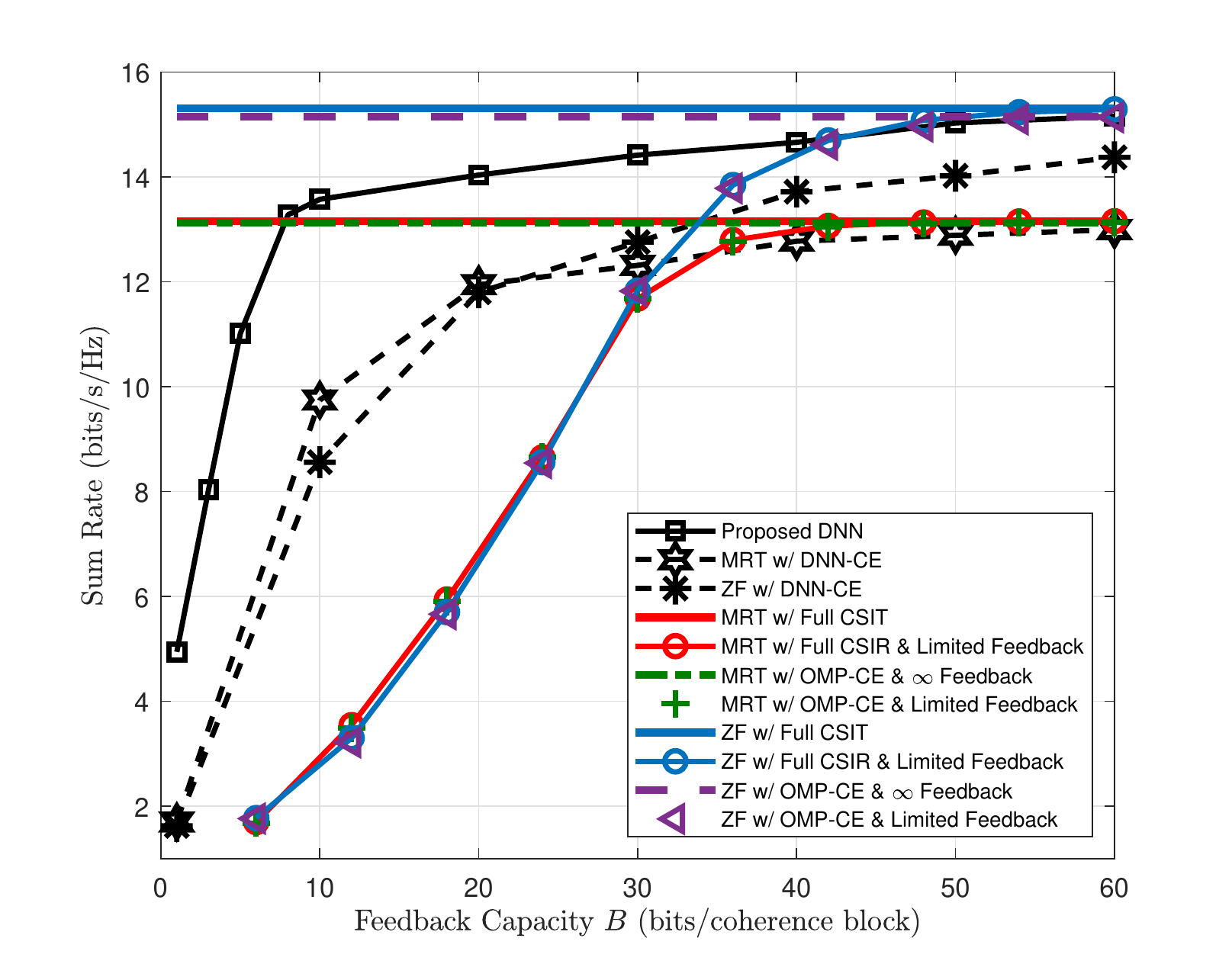}
        \caption{Sum rate achieved by different methods in a 2-user FDD system with $M=64$ and $L =64$.}
       \label{fig:L64}
\end{figure}

Finally, to show that the proposed scheme can eventually approach the performance of ZF with perfect CSIT for sufficiently large $B$ and $L$, in the next experiment, we consider the pilot length $L=M=64$ in which a near perfect channel recovery at the user side may be possible. Fig.~\ref{fig:L64} shows that the proposed method always achieves a higher sum rate as compared to the other limited-feedback baseline methods, while it approaches the performance of the ZF with perfect CSIT at feedback rate about $B\geq50$.

%%%%%%%%%%%%%%%%%%%%%%%%%%%%%%%%%%%%%%
% VI-C) Towards Generalizability w.r.t. $L_p$
%%%%%%%%%%%%%%%%%%%%%%%%%%%%%%%%%%%%%%
\subsection{Towards Generalizability in $L_p$}

In the previous subsection, we train and test the proposed neural network for the same limited-scattering environment, in which the number of channel paths is assumed to be $L_p=2$. In this subsection, we are interested \changeNEW{in numerically investigating} the following two questions. (i) How does the proposed neural network perform if there is a mismatch between the number of paths in the channel realizations of the training data set and that of the test data set? (ii) Can we enhance the robustness of the proposed neural network such that it can perform well for a wide range of channel distributions, i.e., different values of $L_p$? 

%%%%%%%%%%%%%%%%%%%%%%% Fig7 ---> Fig6
\begin{figure}[t]
        \centering
        \includegraphics[width=0.5\textwidth]{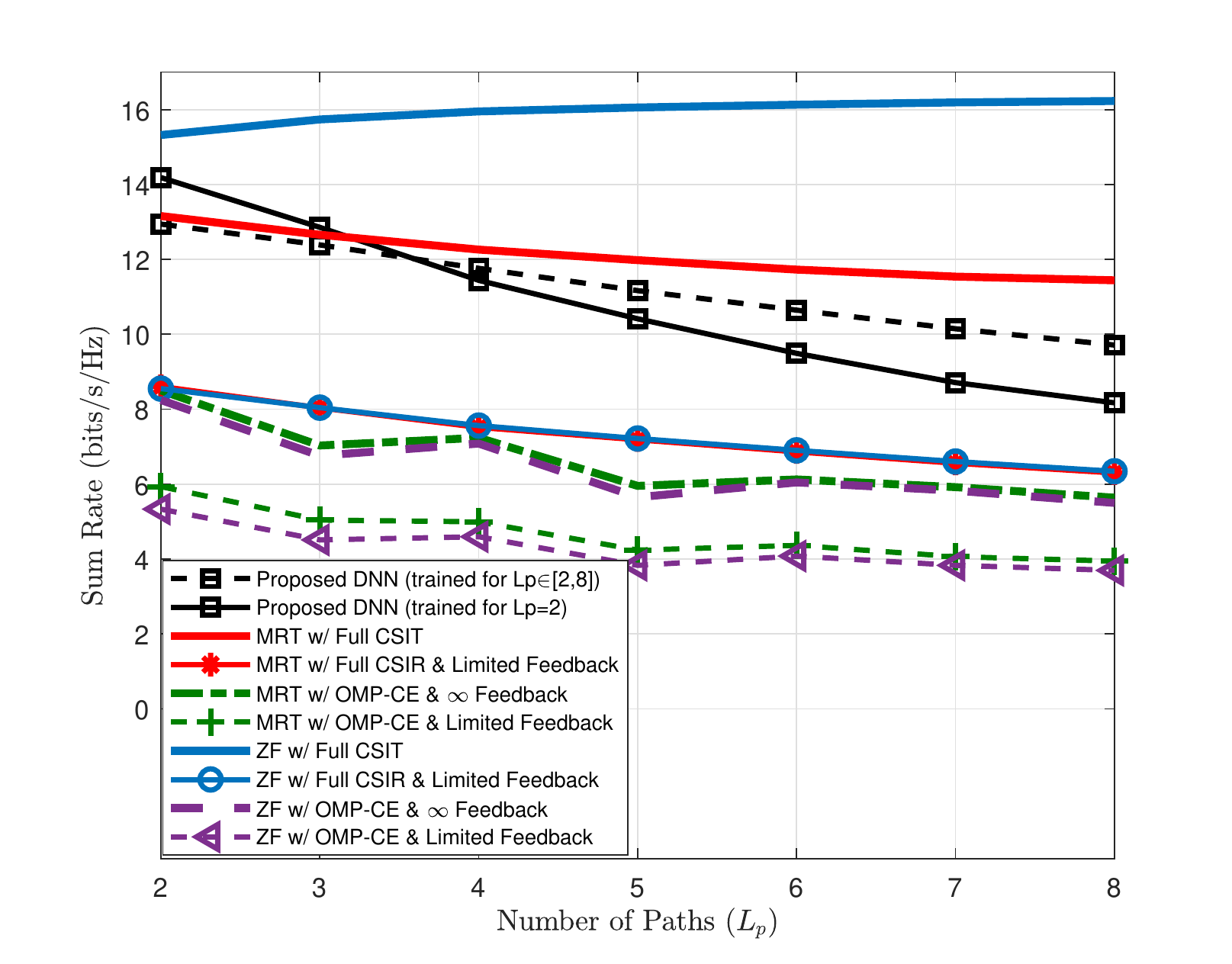}
        \caption{Sum rate achieved by different methods in a 2-user FDD system with $M=64$, $B=30$, and $L=8$.}
       \label{fig:L8_GenLp}
\end{figure}

\begin{figure}[t]
        \centering
        \includegraphics[width=0.5\textwidth]{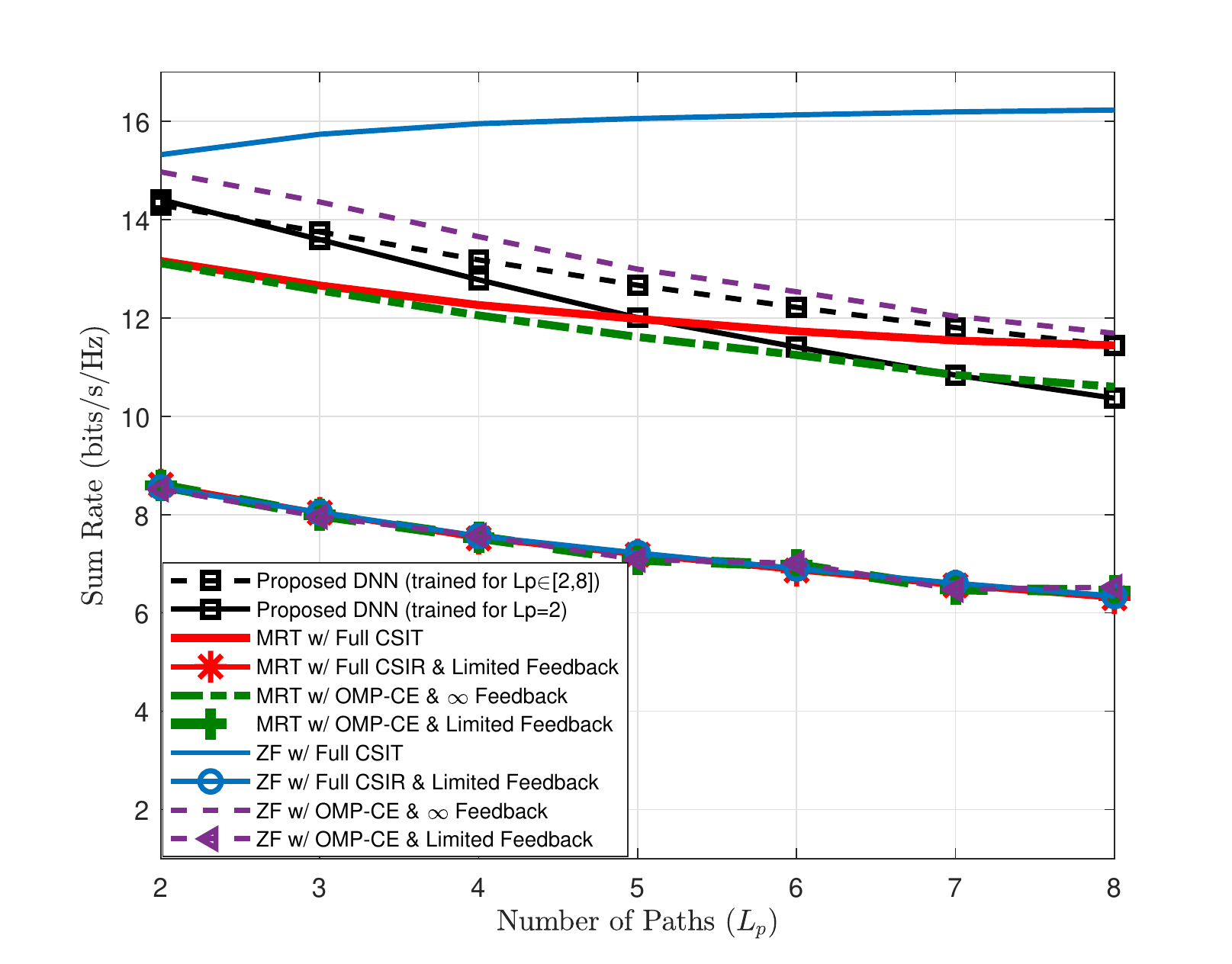}
        \caption{Sum rate achieved by different methods in a 2-user FDD system with $M=64$, $B=30$, and $L=64$.}
       \label{fig:L64_GenLp}
\end{figure}

To numerically investigate the answers for the above questions, we consider a downlink FDD system with $B=30$ while the other parameters remain the same as in the previous experiments. For the proposed method, we train two instances of the proposed neural network; one of them is trained over the channel realizations with $L_p=2$, and the other one is trained using samples with $L_p \in \{2,3,\ldots,8\}$. For the baselines with imperfect CSIT, we assume that the sparse channel parameters of the two strongest paths are estimated at the users and are fed back to the BS. The achievable \changeNEW{sum rates} against the number of scattering paths $L_p$ for $L=8$ and $L=64$ are respectively illustrated in Fig.~\ref{fig:L8_GenLp} and Fig.~\ref{fig:L64_GenLp}. As it can be seen from these two figures, the performance of the DNN that is trained only with $L_p=2$ samples is degraded when it is tested against other values of $L_p$. This is because there is a mismatch between the distribution of the training data set and that of the test data set. This performance degradation is more severe when the training pilot sequence is short, e.g., $L=8$. This is most probably because when the training resources are limited, the proposed DNN does its best to fully exploit the distribution of the input and to tailor the feedback scheme to that particular distribution. However, when the number of training pilot sequences is not the bottleneck, e.g., $L=64$, the trained DNN can potentially handle a wider range of the channel distributions. 

From Fig.~\ref{fig:L8_GenLp} and Fig.~\ref{fig:L64_GenLp}, we also see that the proposed DNN can achieve a more robust performance when it is trained with $L_p \in \{2,3,\ldots,8\}$ samples. This suggests that training the DNN on a wider range of channel parameters can help design more robust systems when perfect prior knowledge about those parameters are not available. Nevertheless, Fig.~\ref{fig:L8_GenLp} and Fig.~\ref{fig:L64_GenLp} show that the proposed DNN still significantly outperforms the other baselines with limited feedback, even when there is a mismatch between the channel parameters in the training set and those in the test set. 

%%%%%%%%%%%%%%%%%%%%%%% Fig8 ----> Fig7

%%%%%%%%%%%%%%%%%%%%%%%%%%%%%%%%%%%%%%
% VI-D) Towards Generalizability in $B$
%%%%%%%%%%%%%%%%%%%%%%%%%%%%%%%%%%%%%%
\subsection{Towards Generalizability in $B$}

\begin{figure}[t]
        \centering
        \includegraphics[width=0.5\textwidth]{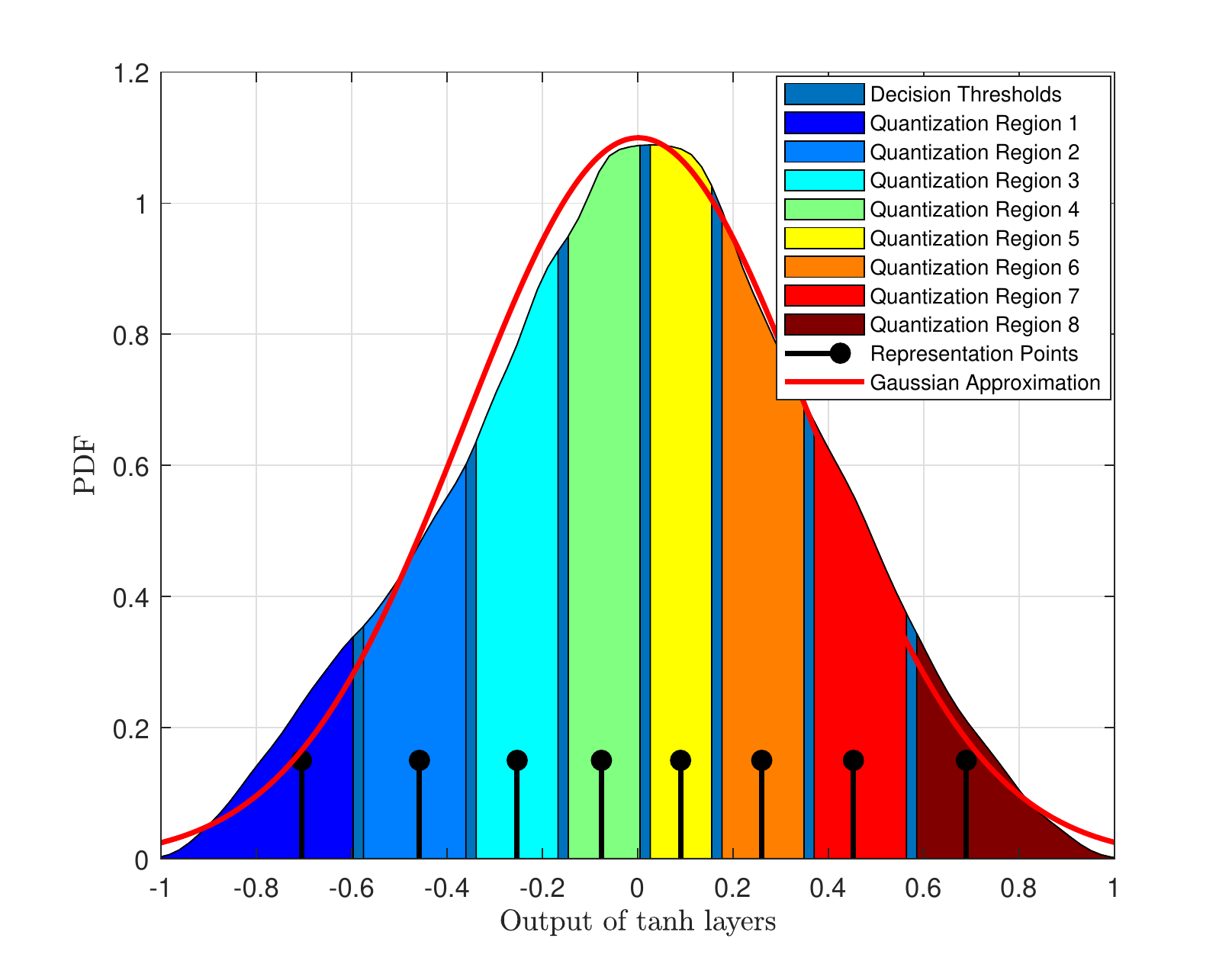}
        \caption{The empirical PDF of the tanh layer output in a proposed modified neural network which has been trained for a system with $M=64$, $K=2$, and $L=8$. This figure also indicates the quantization regions and the corresponding representation points for the optimal $3$-bit quantizer.}
       \label{fig:dist_tanh}
\end{figure}

\begin{figure}[t]
        \centering
        \includegraphics[width=0.5\textwidth]{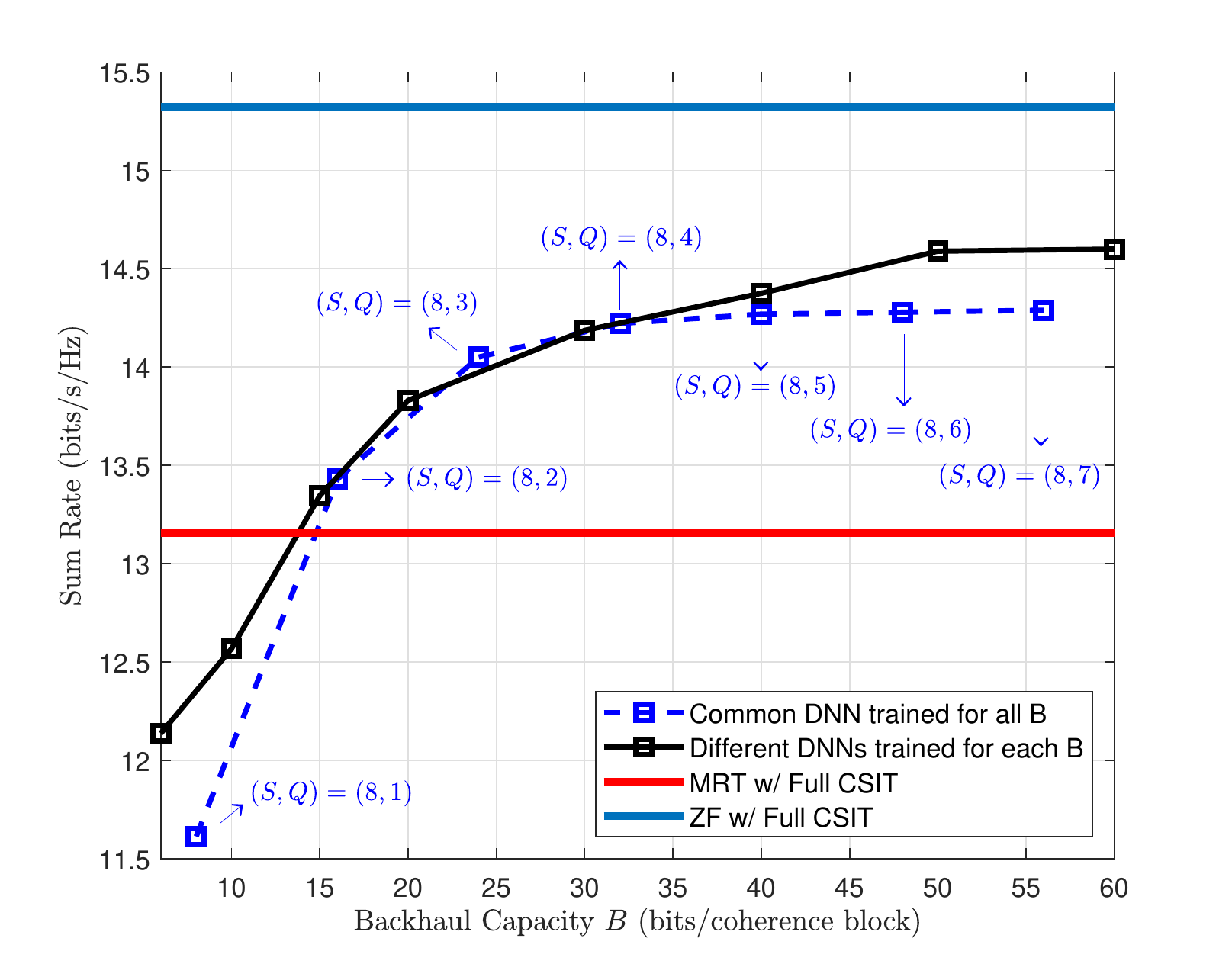}
        \caption{Sum rate achieved by different methods in a 2-user FDD system with $M=64$ and $L =8$.}
       \label{fig:L8_GenB}
\end{figure}

In Section~\ref{sec:proof_of_concept} we have presented the performance of the proposed deep learning-based approach against the capacity of the feedback links $B$. To do so, we propose to train a different instance of the proposed neural network in Fig.~\ref{fig:DNN} for each value of $B$. In practice, it would be more desirable to train a single DNN which can operate for different $B$'s. Accordingly, we now illustrate the performance of the proposed two-step training approach, presented in Section~\ref{sec:gen_B}, which trains a common DNN for different values of $B$. In this experiment, we consider the same setup as in the experiment of Fig.~\ref{fig:L8}. As explained in Section~\ref{sec:gen_B}, in the first step, we train a modified DNN architecture where the $\operatorname{tanh}$ layer with $S$ soft outputs replaces the binary layer. In Fig.~\ref{fig:dist_tanh}, the empirical PDF of the $\operatorname{tanh}$ layer output of the trained modified network is plotted. Interestingly, we observe that this empirical PDF is well approximated as a Gaussian distribution. Next, we obtain the optimal scalar quantizer for this distribution using the Lloyd-Max method. The optimized quantization regions and their corresponding representation points for $Q=3$ bits are shown in Fig.~\ref{fig:dist_tanh}. Finally, we train the parameters of the BS-side DNN while fixing the user-side DNN. The performance of this two-step training approach is shown in Fig.~\ref{fig:L8_GenB}. It can be seen that there is only marginal performance degradation in adopting this training approach, which provides a common DNN that can handle different values of $B$, as compared to using the earlier approach, which requires to train a separate DNN for each feedback capacity value. This shows that the proposed two-step approach can indeed help improve the generalizability of the proposed neural network with respect to the feedback capacity, with the caveat that only integer values of $S$ can be supported in this two-step training approach.

%%%%%%%%%%%%%%%%%%%%%%% Fig9 ---> Fig8

%%%%%%%%%%%%%%%%%%%%%%% Fig10 ---> Fig9

%%%%%%%%%%%%%%%%%%%%%%%%%%%%%%%%%%%%%%
% VI-D) Towards Generalizability w.r.t. $K$
%%%%%%%%%%%%%%%%%%%%%%%%%%%%%%%%%%%%%%
\subsection{Towards Generalizability in $K$}
In all previous experiments, we have evaluated the performance of the proposed DNN for $K=2$. In this last experiment, we investigate whether or not the proposed approach can handle the scenarios in which the BS serves more users. In this experiment, we consider a system with downlink training resources of $L=8$ and $B=30$. Further, to show that the proposed method can handle a larger range of AoDs, in this experiment we model the AoDs as $\theta_{\ell,k} \sim \mathcal{U}(-60^\circ,+60^\circ)$. The other system parameters remain the same as in the previous subsections. We remark that since the input dimension of the decoding DNN is $KB$, for larger values of $K$ we need to increase the capacity of the BS's DNN in order to fully process the input signals. In the simulations, we still employ a $4$-layer DNN at the BS but this time with $[\ell_1^\prime,\ell_2^\prime,\ell_3^\prime,\ell_4^\prime]=[2048,1024,512,2MK]$ number of neurons per layer. 

\begin{figure}[t]
        \centering
        \includegraphics[width=0.5\textwidth]{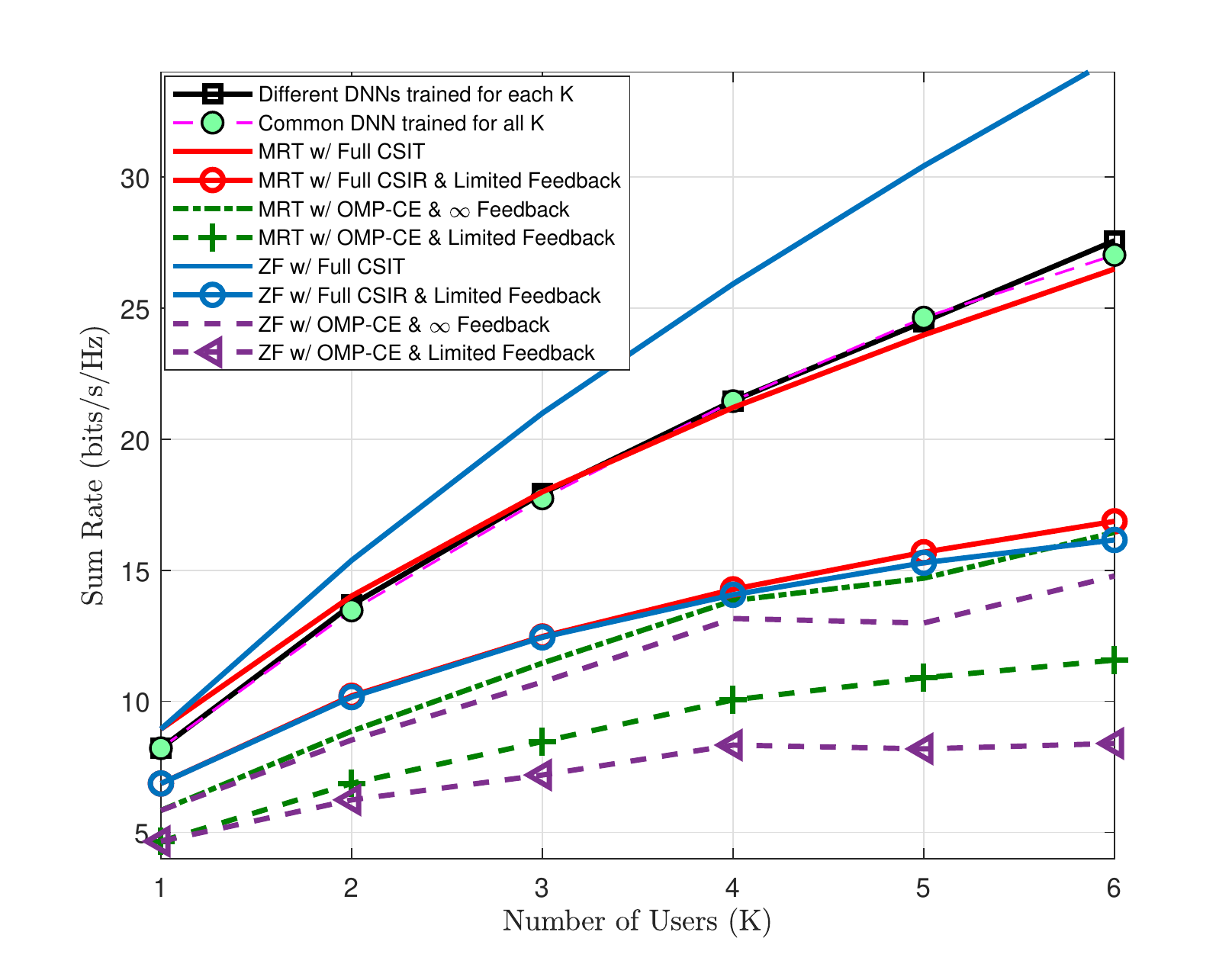}
        \caption{Sum rate achieved by different methods in a $K$-user FDD system with \editb{$M=64$, $B=30$, and $L=8$}.}
       \label{fig:GenK}
\end{figure}

Fig.~\ref{fig:GenK} plots the sum rate against the number of users in the network, i.e., $K$. Fig.~\ref{fig:GenK} shows that similar to the previous experiments with $K=2$, the proposed deep learning framework achieves a much higher rate as compared to all the other \changeNEW{conventional} baselines with limited downlink training resources. However, in the numerical simulations, we observe that for larger values of $K$ the training procedure becomes slower since many parameters have to be jointly designed. To get around with this problem, we also examine the performance of the two-step training procedure proposed in Section~\ref{sec:gen_K}. It can be seen from Fig.~\ref{fig:GenK} that \editb{for this simulation scenario,} the performance of the training approach of Section~\ref{sec:gen_K} is very similar to that of the end-to-end training approach in which all the system parameters are jointly designed. \editb{We remark it is somewhat surprising that a user-side DNN trained for the single-user scenario also works well in the multiuser scenario; but this is found to be true experimentally in this simulation example. %Note that the good performance} of the two-step training approach 
Note that in this example} the different users all have i.i.d.\ channels. Investigating the cases where the channels are correlated or the channel distributions of different users are different can be considered as an interesting direction for future work.

%%%%%%%%%%%%%%%%%%%%%%% Fig11 ---> Fig10

%%%%%%%%%%%%%%%%%%%%%%%%%%%%%%%%%%%%%
%%% VII) Conclusion
%%%%%%%%%%%%%%%%%%%%%%%%%%%%%%%%%%%%%
\section{Conclusion}
\label{sec:conclusion}

This paper proposes a deep learning approach to design a downlink FDD massive
MIMO system with limited feedback which is \changeNEW{formulated as} a DSC problem.  In
particular, this paper presents an end-to-end FDD downlink precoding system,
including the downlink training phase, the uplink feedback phase, and the downlink precoding phase, using a
user-side DNN and a BS-side DNN. We propose a machine learning framework to
jointly design the pilots in the
downlink training phase, the channel estimation and feedback strategy adopted
at the users, and the multiuser precoding scheme at the BS.  This paper also
investigates how to make the proposed DNN architecture more generalizable to
different system parameters. Numerical results show that the proposed DSC
strategy for FDD precoding, which bypasses explicit channel estimation, can
achieve an outstanding performance especially for the scenarios in which the
downlink training pilot length and/or the feedback capacity are very limited. 

%%%%%%%%%%%%%%%%%%%%%%%%%%%%%%%%%%%%%
%%% References
%%%%%%%%%%%%%%%%%%%%%%%%%%%%%%%%%%%%%
\bibliographystyle{IEEEtran}
\bibliography{IEEEabrv,refrenceF}

%%%%%%%%%%%%%%%%%%%%%%%%%%%%%%%%%%%%
%% Biography
%%%%%%%%%%%%%%%%%%%%%%%%%%%%%%%%%%%%
%%%%%%%%%%%%%%%%%% Foad Sohrabi
\begin{IEEEbiography}[{\includegraphics[width=1in,height=1.25in,clip,keepaspectratio]{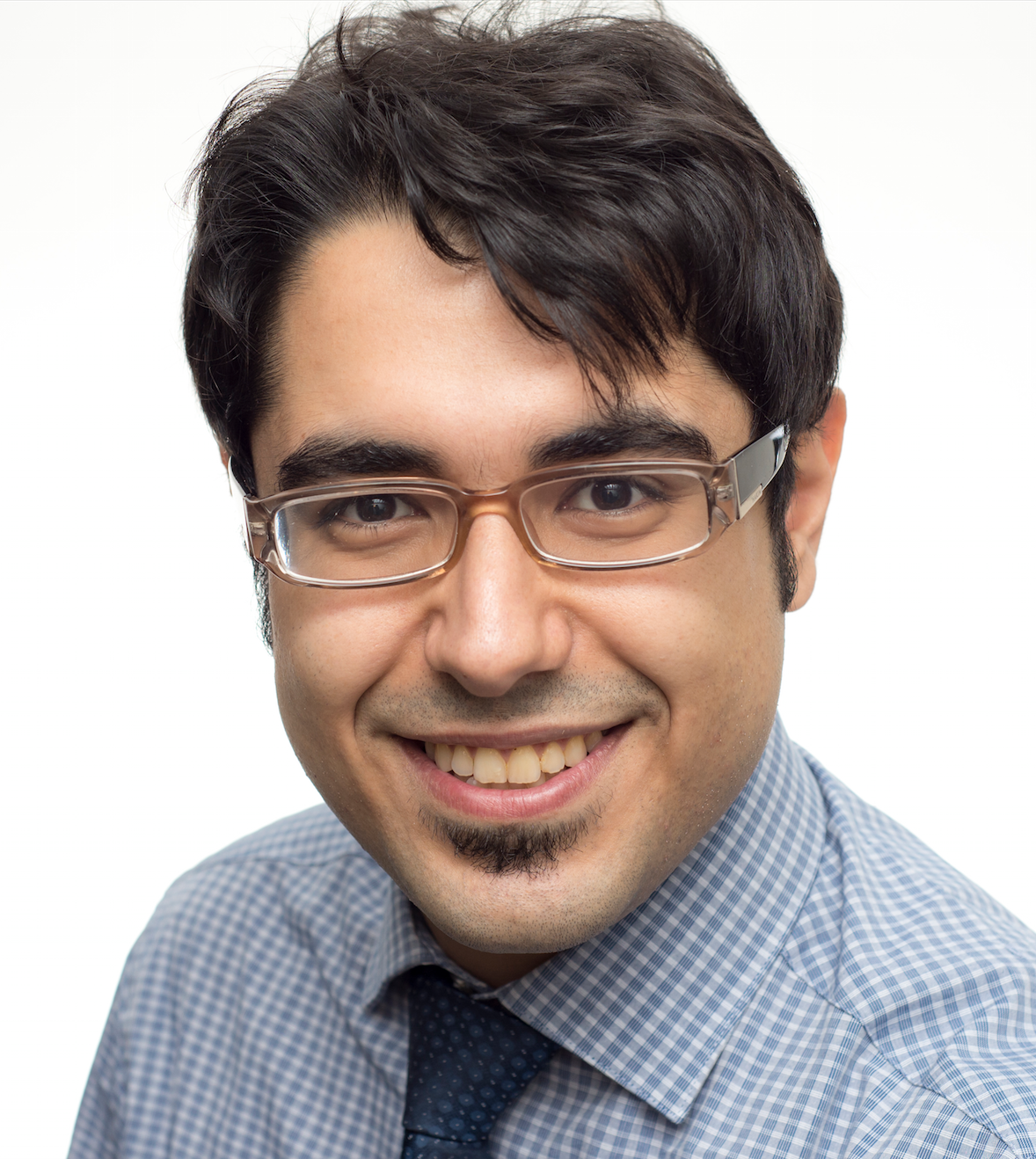}}]{Foad Sohrabi}
(S'13) received the B.A.Sc. degree from the University of Tehran, Tehran, Iran, in 2011, the M.A.Sc. degree from McMaster University, Hamilton, ON, Canada, in 2013, and the Ph.D. degree from the University of Toronto, Toronto, ON, Canada, in 2018, all in electrical and computer engineering. Since 2018, he has been a Post-Doctoral Fellow with the University of Toronto. In 2015, he was a Research Intern with Bell Labs, Alcatel-Lucent, Stuttgart, Germany. His research interests include MIMO communications, optimization theory, wireless communications, signal processing, and machine learning. He was a recipient of the IEEE Signal Processing Society Best Paper Award in 2017.
\end{IEEEbiography}

%%%%%%%%%%%%%%%%%% Kareem Attiah
\begin{IEEEbiography}[{\includegraphics[width=1in,height=1.25in,clip,keepaspectratio]{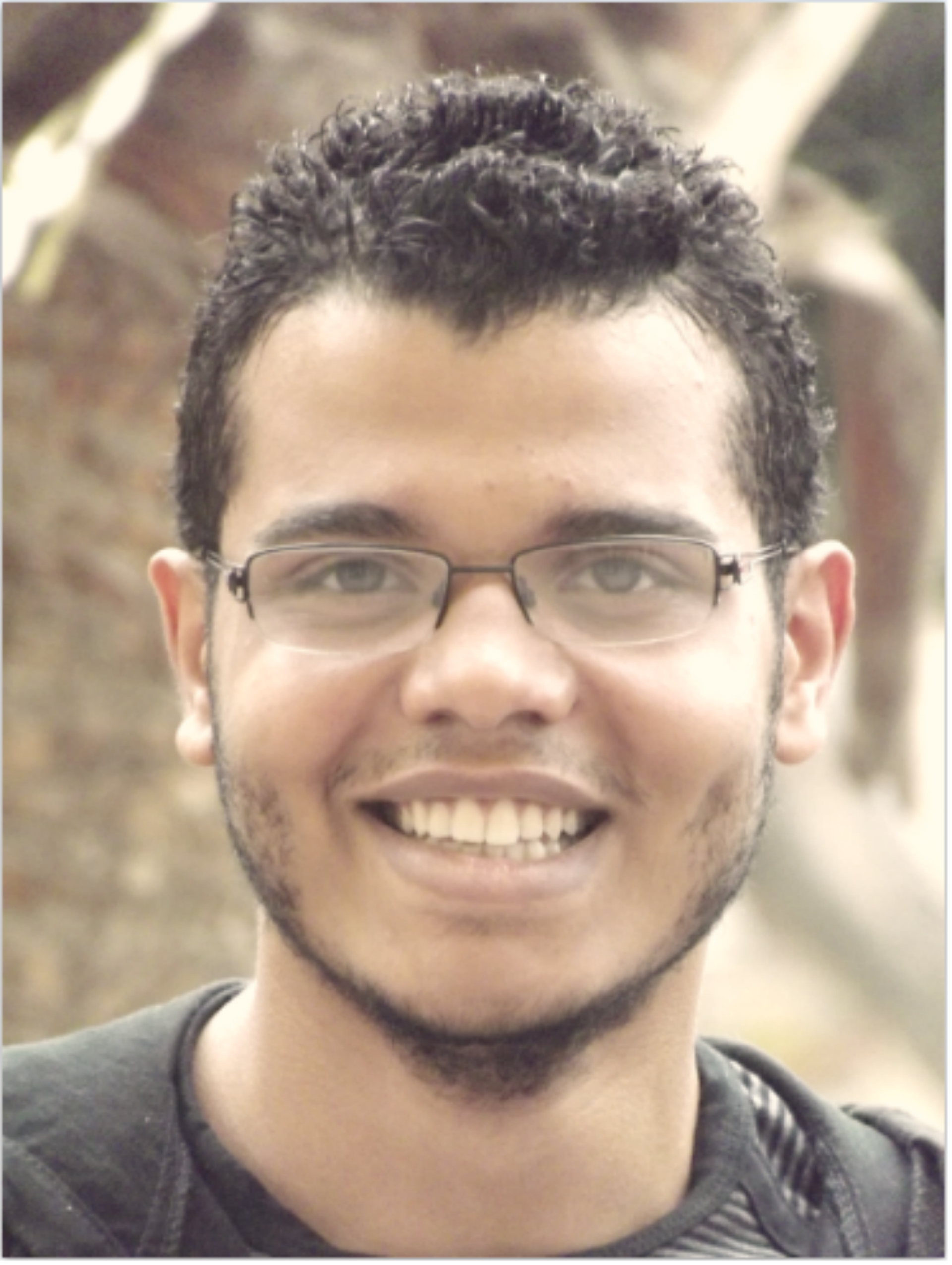}}]
{Kareem M.\ Attiah} received his B.Sc. and M.Sc. degrees in electrical engineering from Alexandria University, Alexandria, Egypt, in 2014 and 2018, respectively. He was a recipient of the 2014 Certificate of Honor from the Egyptian President as the top-ranked student of the Faculty of Engineering, Alexandria University. In the summer of 2014, he was a Visiting Student at Summer@EPFL Internship, École Polytechnique Fédérale de Lausanne (EPFL), Lausanne, Switzerland. In 2017, Mr. Attiah was a member of the Carleton/Ericsson Canada Inc., a collaborative Research Project, and a Visiting Scholar of the Department of System and Computer Engineering, Carleton University. In addition, he is a former Research Assistant with the Communications and Electronics Department, The American University in Cairo (AUC), Cairo, Egypt. As of 2019, he is working towards his Ph.D. degree at the Electrical and Computer Engineering Department, University of Toronto, Toronto, Canada, under the supervision of Prof. Wei Yu. His current research interests include Information Theory, Millimeter-wave Communications, Machine Learning, and Optimization. 
\end{IEEEbiography}

%%%%%%%%%%%%%%%%%%% We YU
\begin{IEEEbiography}[{\includegraphics[width=1in,height=1.25in,clip,keepaspectratio]{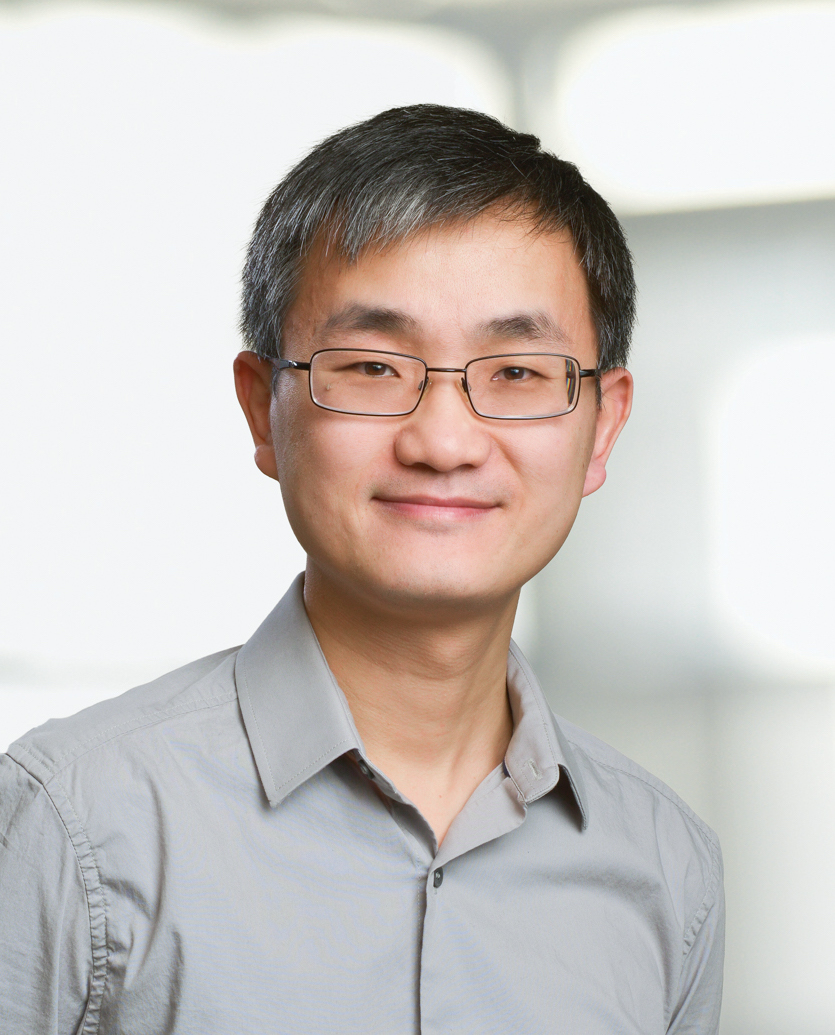}}]
{Wei Yu} (S'97-M'02-SM'08-F'14) received the B.A.Sc. degree in Computer Engineering and Mathematics from the University of Waterloo, Waterloo, Ontario, Canada in 1997 and M.S. and Ph.D. degrees in Electrical Engineering from Stanford University, Stanford, CA, in 1998 and 2002, respectively. Since 2002, he has been with the Electrical and Computer Engineering Department at the University of Toronto, Toronto, Ontario, Canada, where he is now Professor and holds a Canada Research Chair (Tier 1) in Information Theory and Wireless Communications. Prof. Wei Yu is the President of the IEEE Information Theory Society in 2021, and has served on its Board of Governors since 2015. He served as the Chair of the Signal Processing for Communications and Networking Technical Committee of the IEEE Signal Processing Society in 2017-18. Prof. Wei Yu was an IEEE Communications Society Distinguished Lecturer in 2015-16. He is currently an Area Editor for the IEEE Transactions on Wireless Communications, and in the past served as an Associate Editor for IEEE Transactions on Information Theory, IEEE Transactions on Communications, and IEEE Transactions on Wireless Communications. Prof. Wei Yu is a Fellow of the Canadian Academy of Engineering, and a member of the College of New Scholars, Artists and Scientists of the Royal Society of Canada. He received the Steacie Memorial Fellowship in 2015, the IEEE Marconi Prize Paper Award in Wireless Communications in 2019, the IEEE Communications Society Award for Advances in Communication in 2019, the IEEE Signal Processing Society Best Paper Award in 2017 and 2008, the Journal of Communications and Networks Best Paper Award in 2017, and the IEEE Communications Society Best Tutorial Paper Award in 2015. 
\end{IEEEbiography}

\end{document}